\documentclass[12pt,preprint]{aastex}

\def\be{\begin{equation}}
\def\ee{\end{equation}}
\def\ii{\int_0^{+\infty}}
\def\rtr{r_{\rm tr}}
\def\rign{r_{\rm ign}}
\def\rnu{r_\nu}
\def\rout{r_{\rm out}}
\def\rms{r_{\rm ms}}
\def\dM{\dot{M}}
\def\dMign{\dot{M}_{\rm ign}}
\def\dMop{\dot{M}_{\rm opaque}}
\def\dMtr{\dot{M}_{\rm trap}}
\def\Kign{K_{\rm ign}}
\def\Kop{K_{\rm opaque}}
\def\Ktr{K_{\rm trap}}
\def\Fnuc{F_{\rm nuc}}
\def\Fop{F_\nu^{\rm opaque}}
\def\Ftr{F_\nu^{\rm transp}}
\def\Utr{U_\nu^{\rm transp}}
\def\Ubb{U_\nu^{\rm opaque}}
\def\dNop{\dot{N}_\nu^{\rm opaque}}
\def\dNtr{\dot{N}_\nu^{\rm transp}}

\newbox\grsign \setbox\grsign=\hbox{$>$} \newdimen\grdimen \grdimen=\ht\grsign
\newbox\simlessbox \newbox\simgreatbox \newbox\simpropbox
\setbox\simgreatbox=\hbox{\raise.5ex\hbox{$>$}\llap
     {\lower.5ex\hbox{$\sim$}}}\ht1=\grdimen\dp1=0pt
\setbox\simlessbox=\hbox{\raise.5ex\hbox{$<$}\llap
     {\lower.5ex\hbox{$\sim$}}}\ht2=\grdimen\dp2=0pt
\setbox\simpropbox=\hbox{\raise.5ex\hbox{$\propto$}\llap
     {\lower.5ex\hbox{$\sim$}}}\ht2=\grdimen\dp2=0pt
\def\simgt{\mathrel{\copy\simgreatbox}}
\def\simlt{\mathrel{\copy\simlessbox}}

\title{Neutrino-cooled Accretion Disks around Spinning Black Holes}
\usepackage{epsfig}

\author{Wen-xin Chen, Andrei M. Beloborodov\altaffilmark{1}}
\affil{Physics Department and Columbia Astrophysics Laboratory,
Columbia University, 538 West 120th Street New York, NY 10027}

\altaffiltext{1}{Also at Astro-Space Center of Lebedev Physical
Institute, Profsojuznaja 84/32, Moscow 117810, Russia}

\begin{document}

\begin{abstract}
We calculate the structure of accretion disks around Kerr black holes 
for accretion rates $\dot{M}=0.001 - 10 M_\odot$~$s^{-1}$. Such high-$\dot{M}$ 
disks are plausible candidates for the central engine of gamma-ray bursts. 
Our disk model is fully relativistic and treats accurately microphysics of 
the accreting matter: neutrino emissivity, opacity, electron degeneracy, 
and nuclear composition. The neutrino-cooled disk forms above a critical 
accretion rate $\dMign$ that depends on the black hole spin. The disk has 
the ``ignition'' radius $\rign$ where neutrino flux rises dramatically, 
cooling becomes efficient, and the proton-to-nucleon ratio $Y_e$ drops. 
Other characteristic radii are $r_\alpha$ where most of $\alpha$-particles 
are disintegrated, $r_\nu$ where the disk becomes $\nu$-opaque, and $\rtr$ 
where neutrinos get trapped and advected into the black hole.
We find $r_\alpha$, $\rign$, $\rnu$, $\rtr$ and show their dependence on 
$\dot{M}$. We discuss the qualitative picture of accretion and present 
sample numerical models of the disk structure.
All neutrino-cooled disks regulate themselves to a characteristic state
such that: (1) electrons are mildly degenerate, (2) $Y_e\sim 0.1$, 
and (3) neutrons dominate the pressure in the disk.

\end{abstract}



\section{INTRODUCTION}

A tight neutron-star binary loses orbital momentum via gravitational 
radiation and eventually merges, forming a black hole and a transient 
debris disk with a huge accretion rate, comparable to $M_\odot$~$s^{-1}$ 
(see e.g. Ruffert et al. 1997 for numerical simulations). 
A similar accretion disk 
may form inside a collapsing massive star, so called ``collapsar'' 
(Woosley 1993; see McFadyen \& Woosley 1999 for numerical simulations). 
In this case the central parts of the stellar core quickly form a black 
hole of mass $M\sim 2-3 M_\odot$ that grows through accretion on the 
core-collapse timescale $\sim 10$~s.
If the core material has specific angular momentum substantially above 
$GM/c$ an accretion disk forms with $\dot{M}=0.01 - 1M_\odot$~$s^{-1}$.

Accretion disks are known to be efficient producers of relativistic jets 
in various sources associated with black holes, including X-ray binaries
and quasars. It is reasonable to expect that the hyper-accreting disks 
in neutron-star mergers and collapsars produce relativistic jets as well. 
Because of the huge accretion rate, the power of such jets may be enormous. 
If a fraction $\sim 10^{-3}$ of the accretion power $\dot{M}c^2$ is 
channelled to a jet, it will create an explosion with energy 
$\sim 2\times 10^{51}(M_{\rm acc}/M_\odot)$~erg where $M_{\rm acc}$
is the mass accreted through the disk.
Hyper-accreting disk therefore provides a plausible mechanism for powerful
relativistic
explosions observed as gamma-ray bursts (GRBs, see Piran 2004 for a 
review).

Accretion in the disk is driven by viscous stress $t_{r\phi}$ that may be 
expressed as $t_{r\phi}=\alpha p$ where $p$ is pressure inside the disk and
$\alpha<1$ is a dimensionless parameter (Shakura \& Sunyaev 1973).
Viscous stress is created by magnetic fields that are amplified as a 
result of magneto-rotational instability. Numerical simulations
of this instability show that $\alpha=0.01-0.1$ 
(see Balbus \& Hawley 1998 for a review). In this respect, the 
hyper-accreting disk is expected to be similar to normal
accretion disks in X-ray binaries. It is, however, crucially different as 
regards microphysics of the accreting matter and its cooling.
The optical depth 
to photon scattering is enormous and radiation cannot escape;
it is advected by the matter flow into the black hole (or outward by the jet). 
The only possible cooling mechanism is neutrino emission. 
Significant neutrino losses occur at high $\dot{M}>0.01M_\odot$~s$^{-1}$,
and then almost all the accretion energy is carried away by neutrinos.

In any realistic scenario, the black hole has a significant spin as it 
forms from rotating matter and further spun up by accretion.
The spin is likely to help the jet formation through, e.g. Blandford-Znajek 
process. It also increases the overall efficiency of accretion from 6\% 
(zero spin $a=0$) up to 42\% (maximum spin $a=1$). The black-hole spin has 
significant effects on the accretion disk
because it changes dramatically the spacetime metric near the black hole, 
where most of accretion power is released. For example, in the extreme
case of $a=1$, the inner radius 
of the disk is reduced by a factor of 6 compared with the Schwarzschild case.
This leads to a higher temperature and a much higher neutrino intensity.
Therefore, disks around rapidly spinning black holes may create powerful 
jets via neutrino annihilation above the disk.

The structure of neutrino-cooled disks was investigated in a number of
works, and one of them studied accretion onto a spinning Kerr 
black hole (Popham, Woosley, \& Fryer 1999, hereafter PWF). 
PWF used accurate equations of relativistic hydrodynamics in 
Kerr spacetime, however, made simplifying assumptions about the state 
of accreting matter and its neutrino emission, which later turned out 
invalid: (1) Electron degeneracy was neglected. The true degeneracy 
is significant: it strongly suppresses the $e^\pm$ population,
changes the equation of state and neutrino emissivity. 
(2) Neutron-to-proton ratio was not 
calculated and no distinction was made between neutrons and protons 
in the calculation of neutrino emissivity.
The correct ratio is typically $n_n/n_p\sim 10$, which leads to a strong 
suppression of neutrino emission. (3) The produced neutrinos were assumed 
to escape freely. In the case of a Kerr black hole, this assumption breaks 
at $\dot{M}\simgt 0.1M_\odot$~s$^{-1}$ --- then the disk is opaque for 
neutrinos in the inner region where most of accretion power is released.

Recent works discussed one or several of these issues.
For example, Di Matteo, Perna, \& Narayan (2002) focused on the effects 
of neutrino opacity at high $\dot{M}$, however neglected degeneracy and 
assumed $n_n/n_p=1$. 
Kohri \& Mineshige (2002) studied strongly degenerate disks, and
Kohri, Narayan, \& Piran (2005) --- the realistic midly degenerate disks.
None of these works, however, attempted the construction of an accretion 
model in Kerr spacetime --- all assumed a non-rotating black hole
and used a Newtonian or pseudo-Newtonian approximation.
Pruet, Woosley, \& Hoffman (2003) studied the Kerr disk models of PWF
and pointed out that the disk must be neutron rich in the inner region,
however did not attempt to model self-consistently the disk structure.
Beloborodov (2003a) showed that $\beta$-equilibrium is established in disks 
with $\dot{M}\simgt 10^{31}(\alpha/0.1)^{9/5}(M/M_\odot)^{6/5}$~g~s$^{-1}$,
where $M$ is the black hole mass. The relation between temperature, 
density, and neutron richness in equilibrium was studied in that work, 
and the disk was shown to become neutron rich in the inner region if 
$\dot{M}\simgt 10^{31}(\alpha/0.1)(M/M_\odot)^{2}$~g~s$^{-1}$. The 
structure of the disk was not, however, calculated.

In the present paper we develop a self-consistent, fully relativistic 
model of accretion disks around Kerr black holes. Our study
is limited to steady accretion with constant $\dot{M}$, which is a good 
approximation at radii where the accretion timescale is shorter than the 
evolution timescale of the disk (it may be $1-10$~s, depending on the 
concrete scenario of disk formation). The accretion rate is assumed to be
constant with radius, $\dot{M}(r)=const$, neglecting the fraction of 
$\dot{M}$ that may be lost to a jet.
The model employs the customary approximation of one-dimensional 
hydrodynamics (Shakura \& Sunyaev 1973) where the effects of three-dimensional 
MHD turbulence are described by one viscosity parameter $\alpha$.

In \S~2 we summarize the physics of neutrino-cooled disks,
write down the set of relevant equations, and point out where 
progress is made compared with the previous works.
We pay particular attention to the mechanism of cooling that 
couples to and regulates the state of accreting matter.
The method of solution of the disk equations is described in \S~3,
and the results are presented in \S~4.
Our conclusions are summarized in \S~5.


\section{PHYSICS OF NEUTRINO-COOLED DISKS}

\subsection{Outer Boundary Conditions}

Generally, the size of accretion disk is limited by the maximum angular 
momentum of the flow. We here consider only the region where accretion 
is quasi-steady, i.e. the accretion timescale is smaller than the evolution 
timescale of $\dot{M}$. This sets an effective outer boundary $\rout$ which 
depends on the specific scenario of disk formation. In our models,
we choose $\rout$ sufficiently far (well outside the neutrino ignition 
radius) to cover the whole neutrino-cooled region.

Neutrino cooling is negligible outside the ignition radius $\rign$ and the 
viscously dissipated energy is stored in the accretion flow, which 
makes the disk thick in the outer region.
We choose the disk temperature at $\rout$ so that the energy content 
per baryon approximately equals virial energy $\sim GMm_p/r$ 
as is generally the case in any advective flow (e.g. Narayan \& Yi 1994). 
The flow is assumed to be initially made of alpha particles at the outer
boundary. The calculations will show that the flow completely forgets 
the boundary conditions as it approaches the ignition radius: 
$\alpha$-particles are decomposed into free nucleons, and entropy per 
baryon and 
lepton number are regulated to certain values by neutrino emission. 
This convergence makes the details of advective accretion in the outer 
region unimportant. 

Note also that the thick disk in the outer advective region $r\sim\rout$ 
may produce a strong outflow, so that the accretion rate $\dot{M}$ 
decreases with radius. However, even this fact is not so important 
as long as we are given $\dot{M}$ at $r\simgt\rign$ where the disk 
becomes relatively thin and $\dot{M}$ remains constant.

\subsection{One-dimensional Relativistic Hydrodynamics}

The disk is described by vertically averaged quantities such as density 
$\rho$, temperature $T$, pressure $P$, energy density $U$, electron 
chemical potential $\mu_e$, 
neutrino chemical potential $\mu_\nu$, etc. The disk is axially symmetric 
and steady, so all quantities depend of radius $r$ only and we deal with 
a one-dimensional problem. The difference between radius in cylindrical 
and spherical coordinates is neglected as if the disk were 
geometrically thin. This difference is 
$(r_{sph}-r_{cyl})/r\sim (1/2)(H/r)^2\ll 1$ where $H(r)$ is 
half-thickness of the disk at radius $r$. The thin-disk approximation 
is quite accurate inside the ignition radius where $H/r\simlt 0.1-0.4$, 
and less accurate at $r\sim\rout$ where $H/r\sim 0.7-0.8$.

The vertically-averaged model is designed to describe most of the disk 
material, and it may not describe well the ``skin'' of the disk, 
especially if viscous heating is not uniform in the vertical direction. 
The vertical structure of the disk is governed by the unknown vertical 
distribution of heating and the vertical neutrino transport (Sawyer 2003). 
We do not study the vertical structure in this paper and use the simple 
``one-zone'' approximation of the vertically-averaged disk.

The hydrodynamic equations of a relativistic disk express conservation of 
baryon number, energy, and momentum (angular and radial) in Kerr spacetime 
(see Beloborodov 1999 for a review).
Hereafter we use Boyer-Lindquist coordinates $x^\alpha=(t,r,\theta,\phi)$
with the corresponding Kerr metric $g_{\alpha\beta}$, which has two 
parameters: the black hole mass $M$ and its dimensionless spin 
parameter $0<a<1$ (see e.g. Misner, Thorne, \& Wheeler 1973). The disk 
is in the equatorial plane $\theta=\pi/2$ and the fluid motion is 
described by the four-velocity $u^\alpha=dx^\alpha/d\tau=(u^t,u^r,0,u^\phi)$,
where $\tau$ is the proper time of the fluid.
The equation of baryon conservation reads
\be  
\label{eq:baryon}
   u^r=-\frac{\dot M}{4\pi r H\rho},
\ee
where the half thickness $H$ is to be determined below by the equation of 
hydrostatic balance.

The exact differential equations of azimuthal and radial motion may be 
replaced by much simpler and sufficiently accurate conditions following 
Shakura \& Sunyaev (1973) and Page \& Thorne (1974). This ``thin-disk''
approximation neglects terms $\sim (H/r)^2$ and assumes that the fluid is
in Keplerian rotation with angular velocity,
\be
\label{eq:Omega}
 \Omega=\frac{u^\phi}{u^t}=\Omega_{\rm K}
  =\left[\left(\frac{r^3}{GM}\right)^{1/2}+a\,\frac{GM}{c^3}\right]^{-1}.
\ee
A small radial velocity is superimposed on this rotation,
\be
\label{eq:ur}
 u^r=-\frac{\alpha}{S}\,c_s\,\left(\frac{H}{r}\right),
\ee
where $c_s=(P/\rho)^{1/2}$ is the isothermal sound speed and $S(r)$ is 
a numerical factor. This factor is determined by the inner boundary 
condition (zero torque at the last stable orbit $\rms$) and the 
Kerr metric; it is calculated in Appendix~A.
The accuracy of the thin-disk approximation is not perfect
at large radii $r\sim\rout$, where the disk is thick. However, the 
details of the outer region have no effect on the solution for the 
neutrino-cooled disk.

The equation of energy conservation reads
\be  
\label{eq:energy}
  F^{+}-F^{-} = u^r \left[ 
    \frac{d(UH)}{dr}-\frac{(U+P)}{\rho}\frac{d(\rho H)}{dr} \right],  
\ee
where $F^+$ and $F^-$ are the rates of viscous heating and 
cooling per unit area of the disk, as measured by the local observer
corotating with the disk.
$F^-$ depends on the state of the disk matter that will be discussed 
in \S~2.3 below. $F^+$ may be expressed in terms of $\Omega$
and the kinematic viscosity coefficient $\nu$,
\be  
\label{eq:Fplus}
  F^{+}=\nu\,H\,(U+P)\, g^{rr}g_{\phi\phi}(-g^{tt})\gamma^4
                   \left( \frac{d\Omega}{dr} \right) ^2, 
\ee
where $\gamma=(-g^{tt})^{-1/2}u^t$ is the Lorentz factor of fluid
measured in the frame of a local observer with zero angular momentum.
The viscosity coefficient is related to the disk thickness $H$, sound speed
$c_s$, and $\alpha$ by $\nu=(2/3)\alpha c_s H$. 
 
The model described by equations~(\ref{eq:Omega}), (\ref{eq:ur}),
and (\label(\ref{eq:energy}) neglects the effect of the stored energy 
and pressure on the radial and azimuthal dynamics of the disk, and 
retains the advected energy in the energy balance. This approximation is 
reasonable as may be seen from exact hydrodynamical models:
the deviation from Keplerian rotation remains small ($\simlt 10$\%) 
even when the advection effect dominates in the energy balance
(Beloborodov 1998).\footnote{A strong reduction of $\Omega$ below 
$\Omega_{\rm K}$ may occur in the limit of a large disk with no cooling.
This limit does not apply to GRB disks.}  
Then the main effect of advection is the simple 
storage of energy that is not radiated away ($F^+-F^-\neq 0$ in 
eq.~\ref{eq:energy}). It strongly influences pressure and the scale-height 
of the disk, and hence changes $u^r$ according to equation~(\ref{eq:ur}). 
Equations~(\ref{eq:Omega}) and (\ref{eq:ur}) remain, however, good 
approximations.

The vertical balance is given by
\begin{equation}  
\label{eq:vb}
\left( \frac{H}{r} \right) ^{-2}=\frac{J\,GM\rho}{r\,P},
\end{equation} 
where 
\be 
  J=\frac{2(r^2-a r_g\sqrt{2r_gr}
    +0.75a^2 r_g^2)}{2r^2-3r_gr+a r_g\sqrt{2r_gr}}
\ee 
is the relativistic correction to the tidal force in the Kerr metric,
and
\be
  r_g=\frac{2GM}{c^2}.
\ee

In contrast to accretion disks in X-ray binaries and AGN, there is 
one more conservation law that must be taken into account in GRB disks.
The lepton number (or, equivalently, the proton-to-baryon ratio $Y_e$) 
may change with radius because the neutrino and anti-neutrino fluxes 
from the disk may not be equal. The conservation of lepton number is
expressed by the equation,
\be  
\label{eq:Ye}
    \frac{1}{H}(\dot N_{\bar\nu}-\dot N_{\nu})
     =u^r\left[ \frac{\rho}{m_p}\frac{dY_e}{dr}+\frac{d}{dr}
      (n_\nu -n_{\bar\nu}) \right],
\ee
where $\dot N_{\nu}$ and $\dot N_{\bar\nu}$ are the number fluxes of 
neutrinos and anti-neutrinos per unit area (from one face of the disk), 
$n_\nu$ and $n_{\bar\nu}$ are the number densities of neutrinos and 
anti-neutrinos inside the disk; they become significant when the disk 
is opaque (see \S~2.4 below). $Y_e$ is related to the 
neutron-to-proton ratio by $Y_e=(n_n/n_p+1)^{-1}$.

\subsection{Microphysics and Thermodynamic Quantities}

The disk is made of neutrons, protons, $\alpha$-particles, electrons, 
positrons, photons, neutrinos, and anti-neutrinos. The effect of magnetic
field on the particle distribution functions may be neglected 
(see Beloborodov 2003a). The total pressure and energy density are given by
\be
\label{eq:P}
  P=P_b+P_\gamma+P_{e^-}+P_{e^+}+P_\nu+P_{\bar\nu},
\ee
\be
\label{eq:U}
  U=U_b+U_\gamma+U_{e^-}+U_{e^+}+U_\nu+U_{\bar\nu}.
\ee
Here, the baryon pressure and energy density are 
\be
  P_b=\frac{\rho}{m_p}k_BT\left(X_f+\frac{1-X_f}{4}\right),
  \qquad U_b=\frac{3}{2}\,P_b,
\ee
where $X_f$ is the mass fraction of free nucleons and $1-X_f$ is the 
mass fraction of $\alpha$-particles. $X_f$ is found from the equation 
of nuclear statistical equilibrium (see e.g. Meyer 1994),
\be
  4.9\times 10^2\rho_{10}^{-3/2}T_{10}^{9/4}
    \exp\left(-\frac{16.4}{T_{10}}\right)=
    4\left[Y_e-\frac{(1-X_f)}{2}\right]
  \left[1-Y_e-\frac{(1-X_f)}{2}\right](1-X_f)^{-1/2},
\ee
where $\rho_{10}=\rho/10^{10}$~g~cm$^{-3}$ and $T_{10}=T/10^{10}$~K.

The radiation pressure and energy density are
\be
  P_\gamma=\frac{a_rT^4}{3}, \qquad U_\gamma=3P_\gamma,
\ee
where $a_r=7.56\times 10^{-15}$~erg~cm~$^{-3}$~K$^{-4}$ 
is the radiation constant.

Electrons are neither non-degenerate nor strongly degenerate, and
they are not ultra-relativistic at all radii. Therefore, no
asymptotic expansions are valid, and the thermodynamics quantities for
$e^\pm$ must be calculated using the exact Fermi-Dirac distribution. 
The $e^\pm$ pressure and energy density are given by the integrals,
\be
  P_{e^\pm}=\frac{1}{3}\frac{(m_e c)^3}{\pi^2\hbar^3}
     m_e c^2 \ii f(\sqrt{p^2+1},\mp\eta_e) \frac{p^4}{\sqrt{p^2+1}}\,dp,
\ee
\be
  U_{e^\pm}=\frac{(m_e c)^3}{\pi^2\hbar^3} m_e c^2
      \ii f(\sqrt{p^2+1},\mp\eta_e) \sqrt{p^2+1} p^2 \,dp,
\ee
where $f(E,\eta)$ is the Fermi-Dirac distribution function,
\be  
\label{eq:Fermi}
   f(E,\eta)=\frac{1}{e^{\frac{E}{\theta}-\eta}+1},  
\ee 
$\theta=kT/m_ec^2$, $\eta=\mu/kT$ is the dimensionless degeneracy 
parameter, and $\mu$ is chemical potential.
Since $e^\pm$ are in equilibrium with radiation due to fast reactions 
$e^++e^-\leftrightarrow \gamma+\gamma$, and photons have zero chemical 
potential, one has the relation $\mu_{e^-}+\mu_{e^+}=0$. We denote
$\eta_{e^-}$ by $\eta_e$ and use $\eta_{e^+}=-\eta_e$ in 
equations~(\ref{eq:P}) and (\ref{eq:U}). The $e^\pm$ population is 
then completely described by two parameters: $\theta$ and $\eta_e$.

The number densities of $e^-$ and $e^+$ are 
\be  
   n_{e^\pm}=\frac{(m_e c)^3}{\pi^2\hbar^3} 
        \ii f(\sqrt{p^2+1},\mp\eta_e) p^2\,dp.  
\ee 
The disk matter is neutral, which implies 
\be  
\label{eq:charge}
  n_{e^-}-n_{e^+}=Y_e \frac{\rho}{m_p}. 
\ee 
This gives a relation between $\theta$, $\eta_e$, $\rho$, and $Y_e$.

The contribution of $\nu$ and ${\bar\nu}$ to $P$ and $U$ becomes 
noticeable only in very opaque disks where neutrinos are completely 
thermalized and described by Fermi-Dirac distributions with chemical 
potentials $\mu_\nu$ and $\mu_{\bar\nu}=-\mu_\nu$ (see \S~\ref{sc:opaque}). 
$U_\nu$ and $U_{\bar\nu}$ in the opaque disk are given by 
equations~(\ref{eq:Unu}) and (\ref{eq:Unubar}) below. The corresponding 
pressures are $P_\nu=U_\nu/3$ and $P_{\bar\nu}=U_{\bar\nu}/3$.

\subsection{Cooling}

The cooling of the disk $F^-$ may be written as a sum of three terms,
\be
  F^-=\Fnuc+F_\nu+F_{\bar\nu}.
\ee
Here $\Fnuc$ describes the consumption of heat by the disintegration
of $\alpha$ particles as the flow approaches the black hole,
\be
  \Fnuc=6.8\times 10^{28}\rho_{10}\frac{dX_f}{dr}\,u^r\,H,
\ee
where all quantities are expressed in cgs units.
This cooling is dominant in an extended region around $100r_g$ 
where $\alpha$-particles gradually disintegrate (see \S~4 below).

The terms $F_\nu$ and $F_{\bar\nu}$ represent the cooling due
to emission of neutrinos and anti-neutrinos. 
There are four different channels of neutrino emission 
(see e.g. Kohri \& Mineshige 2002):
(1) electron capture onto protons $p+e^{-}\rightarrow n+\nu$ and positron 
capture onto neutrons $n+e^{+}\rightarrow p+\bar\nu$,
(2) pair annihilation $e^{+}+e^{-}\rightarrow \nu +\bar\nu$,
(3) nucleon-nucleon bremsstrahlung $n+n\rightarrow n+n+\nu +\bar\nu$, and
(4) plasmon decay $\tilde\gamma\rightarrow \nu +\bar\nu$. 
The $e^\pm$ capture strongly dominates neutrino emission in GRB disks,
and the other three channels may be safely neglected. The $e^\pm$ 
annihilation into $\nu\bar\nu$
makes a small contribution even when electron degeneracy is neglected 
(PWF). When degeneracy is taken into account, the positron population is 
suppressed and the reaction rate becomes completely negligible.
Bremsstrahlung and plasmon decay are important only at extremely high 
degeneracy and negligible in GRB disks. 

We calculate below $F_\nu$ and $F_{\bar\nu}$ due to the $e^\pm$ capture 
onto nucleons. This calculation is different in the transparent and 
opaque regions of the disk.

\subsubsection{Neutrino Emission from $\nu$-transparent Disk}

If the disk is transparent to the emitted neutrinos and anti-neutrinos, 
the emerging neutrino flux equals the vertically integrated emissivity
that is found, e.g., in Shapiro \& Teukolsky (1983). This gives,
\be  
\label{eq:Ftr}
   F_\nu=H\,Y_e\frac{\rho}{m_p}\,K\,m_e c^2 
      \ii f(E+Q,\eta_e) (E+Q)^2 [1-\frac{1}{(E+Q)^2}]^{1/2} E^3 \,dE,
\ee 
\be  
    F_{\bar\nu}=H\,(1-Y_e)\frac{\rho}{m_p}\,K\,
      m_e c^2 \int_{Q+1}^{+\infty} f(E-Q,-\eta_e) (E-Q)^2 
      \left[1-\frac{1}{(E-Q)^2}\right]^{1/2} E^3 \,dE.
\ee 
Here $Q=(m_n-m_p)/m_e=2.53$, $K=6.5\times 10^{-4}$~s$^{-1}$,
and $f(E,\eta)$ is the Fermi-Dirac distribution (eq.~\ref{eq:Fermi}).

Similarly, one finds the number flux of neutrinos, which will
be used in the equation of lepton number conservation,
\be  
\label{eq:dNtr}
   \dot N_{\nu}=H\,Y_e\frac{\rho}{m_p} \,K\,
 \ii f(E+Q,\eta_e) (E+Q)^2 \left[1-\frac{1}{(E+Q)^2}\right]^{1/2} E^2 \,dE,
\ee 
\be  
   \dot N_{\bar\nu}=H\,(1-Y_e)\frac{\rho}{m_p} 
      \,K\,\int_{Q+1}^{+\infty} f(E-Q,-\eta_e) (E-Q)^2 
        \left[1-\frac{1}{(E-Q)^2}\right]^{1/2} E^2 \,dE. 
\ee 

\subsubsection{Neutrino Emission from $\nu$-opaque Disk}
\label{sc:opaque}

Inside a $\nu$-opaque disk, neutrinos relax to thermal equilibrium:
a detailed balance is established between absorption and emission. 
Then neutrinos are described by the Fermi-Dirac distribution, and the 
energy flux of escaping neutrinos may be written as
\be  
\label{eq:Fop}
   F_\nu=\frac{U_\nu c}{1+\tau_\nu},
\ee
where
\be  
\label{eq:Unu}
  U_\nu=\frac{(m_e c)^3}{2\pi^2\hbar^3} m_e c^2 \ii f(E,\eta_\nu) E^3 \,dE   
\ee 
is the energy density of thermalized neutrinos inside the disk and
$\tau_\nu$ is the total optical depth seen by $\nu$, including absorption
and scattering (the cross sections of relevant processes are given in 
Appendix~B). 
The chemical potential of neutrinos $\eta_\nu=\mu_\nu/kT$ that appears
in equation~(\ref{eq:Unu}) is related to $\mu_e$, $\mu_p$, and $\mu_n$ 
because the detailed equilibrium $\nu+n\leftrightarrow e^-+p$ is established.
The relation $\mu_\nu+\mu_n=\mu_e+\mu_p$ then gives (see Beloborodov 2003a),
\be  
  \eta_e-\eta_\nu=\ln \left(\frac{1-Y_e}{Y_e}\right) + \frac{Q}{\theta}. 
\ee 

When the disk is opaque also for anti-neutrinos, one finds
\be  
   F_{\bar\nu}=\frac{U_{\bar\nu} c}{1+\tau_{\bar\nu}}, \hspace{3pc}
\ee
\be
\label{eq:Unubar}
       U_{\bar\nu}=\frac{(m_e c)^3}{2\pi^2\hbar^3} m_e c^2 
               \ii f(E,-\eta_\nu) E^3 \,dE,
\ee 
where we have used $\eta_{\bar\nu}=-\eta_\nu$.

Finally, the number fluxes of $\nu$ and $\bar{\nu}$ that escape the disk
are given in the opaque regime by 
\be  
\label{eq:dNop}
   \dot N_{\nu}=\frac{n_\nu c}{(1+\tau_\nu)}, \hspace{2pc} 
   n_\nu=\frac{(m_e c)^3}{2\pi^2\hbar^3} \ii f(E,\eta_\nu) E^2 \,dE,  
\ee 
\be  
   \dot N_{\bar\nu}=\frac{n_{\bar\nu} c}{(1+\tau_{\bar\nu})}, \hspace{2pc} 
    n_{\bar\nu}=\frac{(m_e c)^3}{2\pi^2\hbar^3} \ii f(E,-\eta_\nu) E^2 \,dE.
\ee

\subsubsection{Transition Between Transparent and Opaque Regions}
\label{sc:transition}

Two changes happen at the transition to the opaque regime:
(1) the probability of direct neutrino escape is reduced as 
$(1+\tau_\nu)^{-1}$, and (2) neutrinos get thermalized. Note that the 
rates of neutrino scattering and absorption are comparable, and a 
large optical depth $\tau_\nu\gg 1$ implies that neutrinos are completely
reabsorbed in the disk. The neutrino spectrum significantly
changes at the transition: it is described by Fermi-Dirac distribution 
in the opaque region, with a non-zero chemical potential $\mu_\nu$.
In particular, the mean energy of neutrinos, $\bar{E}_\nu/kT$, changes. 

We model the transition using the following approximate method.
First we calculate the energy density of neutrinos that would be obtained in 
the transparent and opaque limits: $\Utr=F_\nu/c$ from equation~(\ref{eq:Ftr}) 
and $\Ubb$ from equation(~\ref{eq:Unu}), and define the parameter
\be
  x=\frac{\Utr}{\Utr+\Ubb}.
\ee
If the disk is transparent we must find $\Utr\ll\Ubb$ (no reabsorption) 
and $x\rightarrow 0$. If it is opaque, $\Utr\gg\Ubb$ (strong reabsorption)
and $x\rightarrow 1$. In our calculations, the neutrino flux is approximated
by
\be
  F_\nu=\left\{ \begin{array}{ll} 
         \Utr c\,(1+\tau_\nu)^{-1} &  \mbox{if $x<\frac{1}{2}$}, \\ 
         \Ubb c\,(1+\tau_\nu)^{-1} &  \mbox{if $x\geq\frac{1}{2}$}.
                \end{array}  
        \right. 
\ee
This expression is continuous at the transition point $x=1/2$.

To model the change of $\dot{N}_\nu$ at the transition we 
approximate the mean energy of the emitted neutrinos by
\be
  \bar{E}_\nu=(1-x)\,\frac{\Ftr}{\dNtr}+x\,\frac{\Fop}{\dNop}.
\ee
Then we define
\be
  \dot{N}_\nu=\frac{F_\nu}{\bar{E}_\nu},
\ee
which smoothly changes across the transition and has the correct limits at 
$x=0$ and $x=1$. Transition to the opaque region for anti-neutrinos is 
calculated in the same way. 

Finally, we note that emission of muon and tau neutrinos is 
orders of magnitude weaker compared with the emission of electron neutrinos
by $e^\pm$ capture reactions.
The main emission mechanism of $\nu_\mu$ and ${\bar\nu}_\mu$ is
nucleon-nucleon bremsstrahlung. The corresponding emissivity when nucleons
are non-degenerate ($\rho< 10^{14}$~g~s$^{-1}$) is given by
(Thompson, Burrows, \& Horvath 2000)
\be
   \dot{q}_{\nu_\mu{\bar\nu}_\mu}\approx 10^{30}\zeta\rho_{14}^2
 \left(\frac{kT}{\rm MeV}\right)^{5.5} \; {\rm erg~cm}^{-3}{\rm~s}^{-1},
\ee 
where $\zeta\sim 0.1-1$ is a numerical factor. At high densities, when 
nucleons become degenerate, $\dot{q}_{\nu_\mu{\bar\nu}_\mu}$ saturates
at about $10^{34}$~erg~cm$^{-3}$~s$^{-1}$. The muon-neutrino emissivity 
will be found to be well below $F^+/H$ and therefore may be neglected.

\subsection{Comparison with Previous Works}

Recent works on neutrino-cooled disks 
assumed a Schwarzschild black hole, and the main 
advantage of our model is that it is fully relativistic and describes
accretion by Kerr black holes. Other advances compared with the most
recent work by Kohri et al. (2005) are as follows. 
(1) Inclusion of the zero-torque boundary condition at the last stable
orbit. This has a strong effect on the heating rate $F^+$ and the radial
velocity $u^r$. The effect of inner boundary is described by the factor 
$S(r)$, which is much smaller than unity in the hottest region of the disk, 
$S\sim 0.1$.
(2) Advection of heat and lepton number is treated accurately, using the
differential equation of radial transport. This is essential since the disk 
is far from being self-similar and no analytical approximations are valid.
The set of equations then becomes more complicated, however, it allows one 
to calculate the global model of the disk, from the outer advective region 
to the last stable orbit. 
(3) The transition to the opaque region is treated accurately. 
We find that there is only one thermalized neutrino species 
(electron neutrino), and its chemical potential $\mu_\nu$ is obtained 
from the detailed equilibrium rather than assumed to be zero. 
(4) A corrected cross-section is used for anti-neutrino absorption by 
protons, which takes into account the proton recoil (see Appendix~B).
This minor refinement is interesting only in high-$\dot{M}$ disks 
that produce high-energy neutrinos.


\section{METHOD OF SOLUTION}

The disk is described by the set of coupled equations most of which are 
local, i.e. relate local parameters at a given radius. Two equations,
however, are differential (eqs.~\ref{eq:energy} and \ref{eq:Ye}); they 
state conservation of energy and lepton number and contain advection terms
that describe radial transport of energy and $Y_e$. We therefore have
to specify two boundary conditions. 

Our outer boundary is outside the neutrino-cooled disk, in the 
advective region, where neutrino emission may be neglected and matter 
is dominated by $\alpha$-particles. So, one of our boundary conditions 
is $Y_e(\rout)=0.5$. 

As the other boundary condition one may specify any parameter that 
gives a reasonable approximation to the advection-dominated 
solution (Narayan, Piran, \& Kumar 2001). For example, one could specify
a certain value of $(H/r)^2\sim 1/2$ that is characteristic for advective 
disks, and calculate all other parameters at $\rout$
using the local equations. There is a limited freedom in the choice of 
$H(\rout)$, which reflects physical uncertainties in the behavior
of the disk outside our boundary $\rout$: how far the disk extends beyond 
$\rout$ and how much mass and energy it has lost to a wind. These 
uncertainties, however, have no impact on the neutrino-cooled disk at 
$r<\rign$ as we verify directly by varying the outer boundary condition:
the solution we get at $r\simlt\rign\ll\rout$ is the same in all cases. 
Instead of $H/r$, one may use energy per baryon $U/\rho$ as a free parameter
at the outer boundary. In the advective disk, $U/\rho$ is comparable to 
the virial specific energy $GM/r$. In the sample models shown below we 
specify $U/\rho=GM/r$ at $r=\rout$.

The set of disk equations is a complicated mixture of differential and
algebraic equations, which involve integrals of the Fermi-Dirac distribution.
We solve these equations on a logarithmic grid $r_i$ ($i=0..N$), starting 
at $r_0=\rout$ and moving inward. At each step, we need to find the 
parameters of the disk at $r=r_i$ using the known parameters at 
$r_{i-1}>r_i$ from the previous step.

The state of matter at any radius is described by $Y_e$ and two independent 
thermodynamic quantities, which we choose to be $\theta=kT/m_ec^2$ and 
$\eta_e=\mu_e/kT$ (because they enter as parameters in the integrals of
Fermi-Dirac distribution). Density $\rho$, pressure $P$, and energy density 
$U$ are expressed in terms of $\theta$, $\eta_e$, and $Y_e$ as explained in 
\S~2.3. For example, the expression for $\rho$ is given by the charge 
neutrality equation~(\ref{eq:charge}).  
The computational problem now reduces to finding $\theta$, $\eta_e$, and 
$Y_e$ at radius $r_i$ given the known parameters at $r_{i-1}$.

$Y_e(r_{i})$ is easily found using the differential equation~(\ref{eq:Ye})
since $Y_e(r_{i-1})$ is known from the previous step. The main difficulty
is in the calculation of $\theta(r_i)$ and $\eta_e(r_i)$. 
As two independent equations for $\theta$ and $\eta_e$ we choose 
hydrostatic balance~(\ref{eq:vb}) and energy equation~(\ref{eq:energy}).
The scale-height $H$ that appears in both equations may be expressed
in terms of $\rho$ using equations~(\ref{eq:baryon}), (\ref{eq:ur}), and
(\ref{eq:vb}),
\be
\label{eq1}
 \left(\frac{H}{r}\right)^3
  =\frac{S}{\alpha}\frac{\dM}{4\pi r^2\rho}\,\left(\frac{r}{GMJ}\right)^{1/2}.
\ee
Then the hydrostatic equation becomes
\be
   P=\left(\frac{S}{\alpha}\frac{GM\dM J}{4\pi r^3}\right)^{2/3}
     \,\rho^{1/3},
\ee
where $r=r_i$. 
Both $P$ and $\rho$ are known functions of $\theta$ and $\eta_e$, 
which are evaluated numerically. This gives one equation for $\theta$ and 
$\eta_e$.

The second (energy) equation is differential and must be discretized,
\be
\label{eq2}
  F^+-F^-=u^r\left[\frac{U_iH_i-U_{i-1}H_{i-1}}{r_i-r_{i-1}}
  -\frac{(U+P)}{\rho}\frac{(\rho_iH_i-\rho_{i-1}H_{i-1})}{r_i-r_{i-1}}\right].
\ee
All quantities taken at point $r_{i-1}$ are known and all quantities taken
at $r_i$ are functions of $\theta$ and $\eta_e$. This gives the second 
equation.

We find $\theta$ and $\eta_e$ that satisfy both equations~(\ref{eq1}) and
(\ref{eq2}) numerically, using a direct search in the $(\theta,\eta_e)$ 
plane. Once $\theta$ and $\eta_e$ are found at $r_i$, we calculate all
parameters of the disk at this radius and move on to the next step 
$r_{i+1}$.


\section{RESULTS}

\begin{figure}
\begin{center}
\plottwo{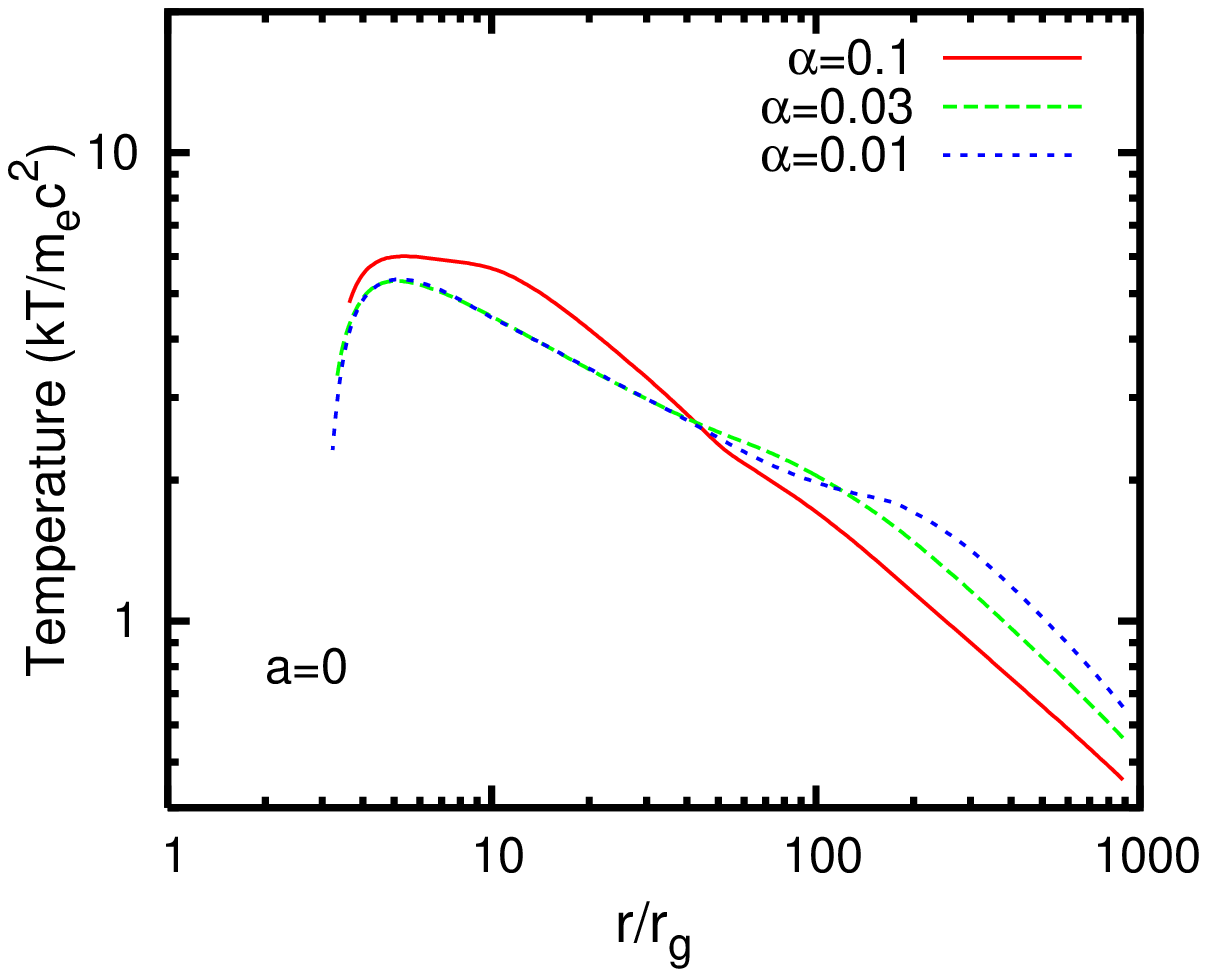}{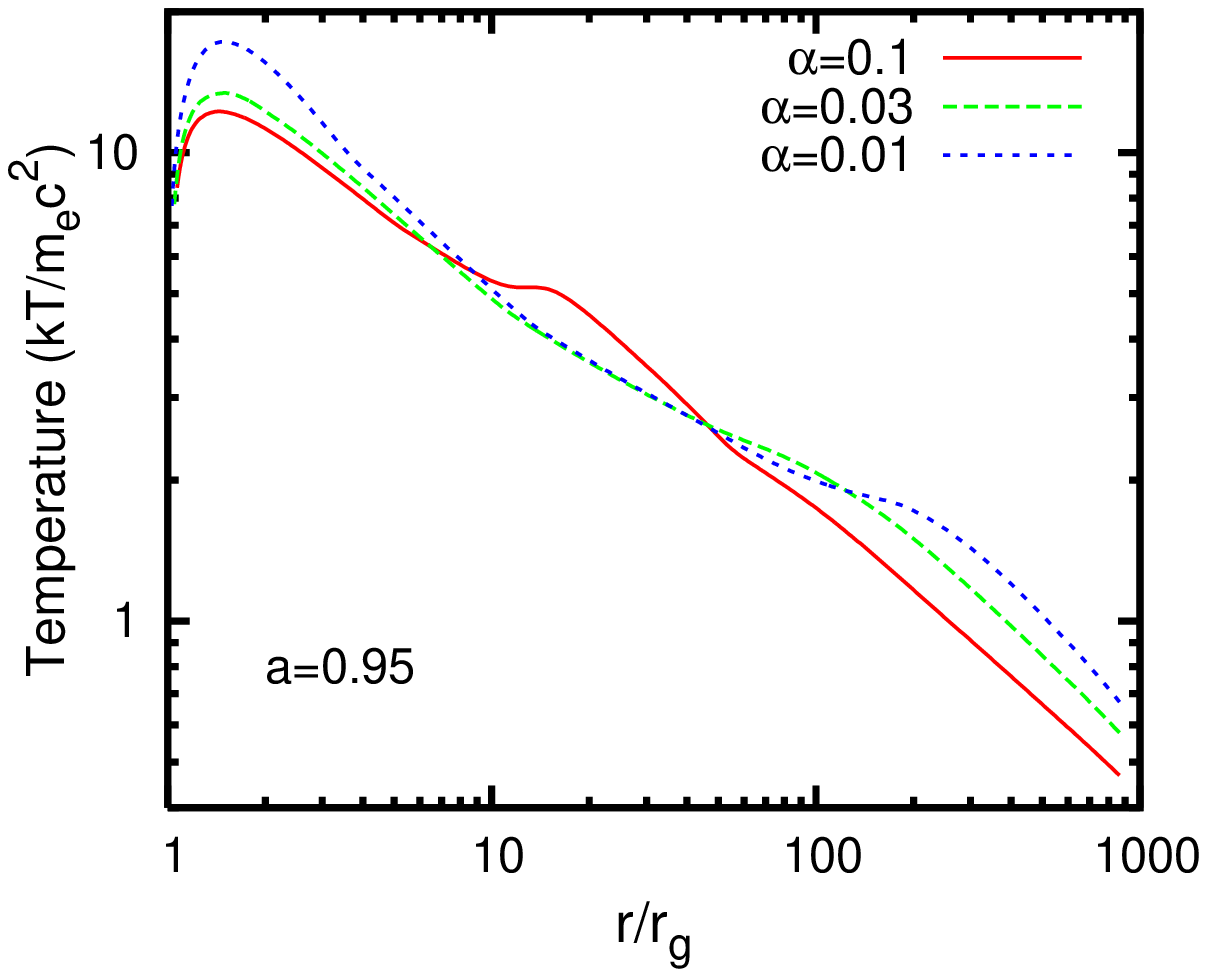}
\label{fig:T}
\caption{Temperature of the accretion disk with $\dM=0.2M_\odot$~s$^{-1}$
around a black hole with mass $M=3M_\odot$. Three disk models are shown 
with different viscosity parameters $\alpha=0.1$, 0.03, and 0.01.
{\it Left panel}: Schwarzschild black hole ($a=0$). 
{\it Right panel:} Kerr black hole ($a=0.95$). Radius is shown in units
of $r_g=2GM/c^2$, and temperature in units of $m_ec^2=0.511$~MeV. }
\end{center}
\end{figure}

\begin{figure}
\begin{center}
\plottwo{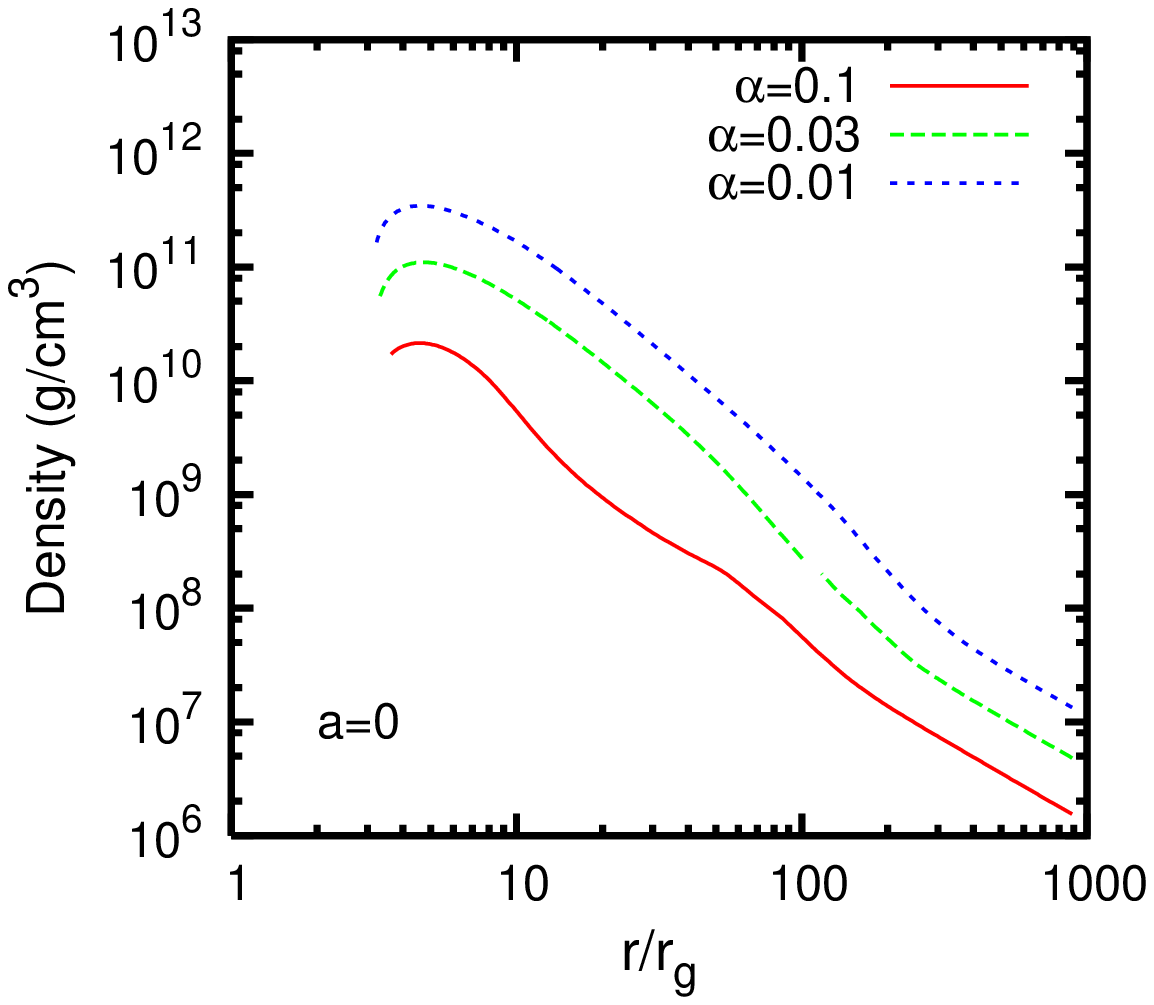}{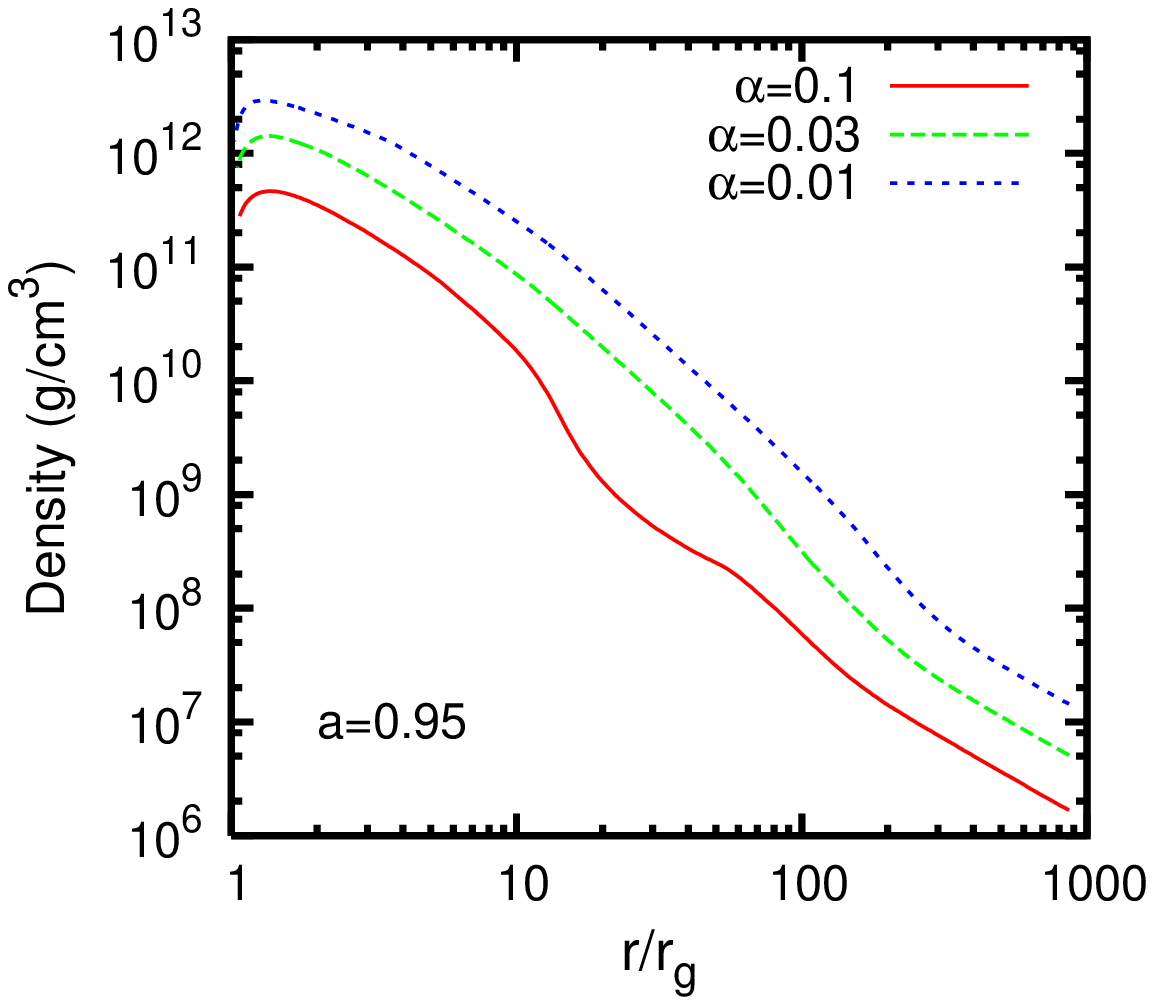}
\label{fig:rho}
\caption{Density $\rho(r)$ for the same disk models as in Fig.~1.}
\end{center}
\end{figure}

\begin{figure}
\begin{center}
\plottwo{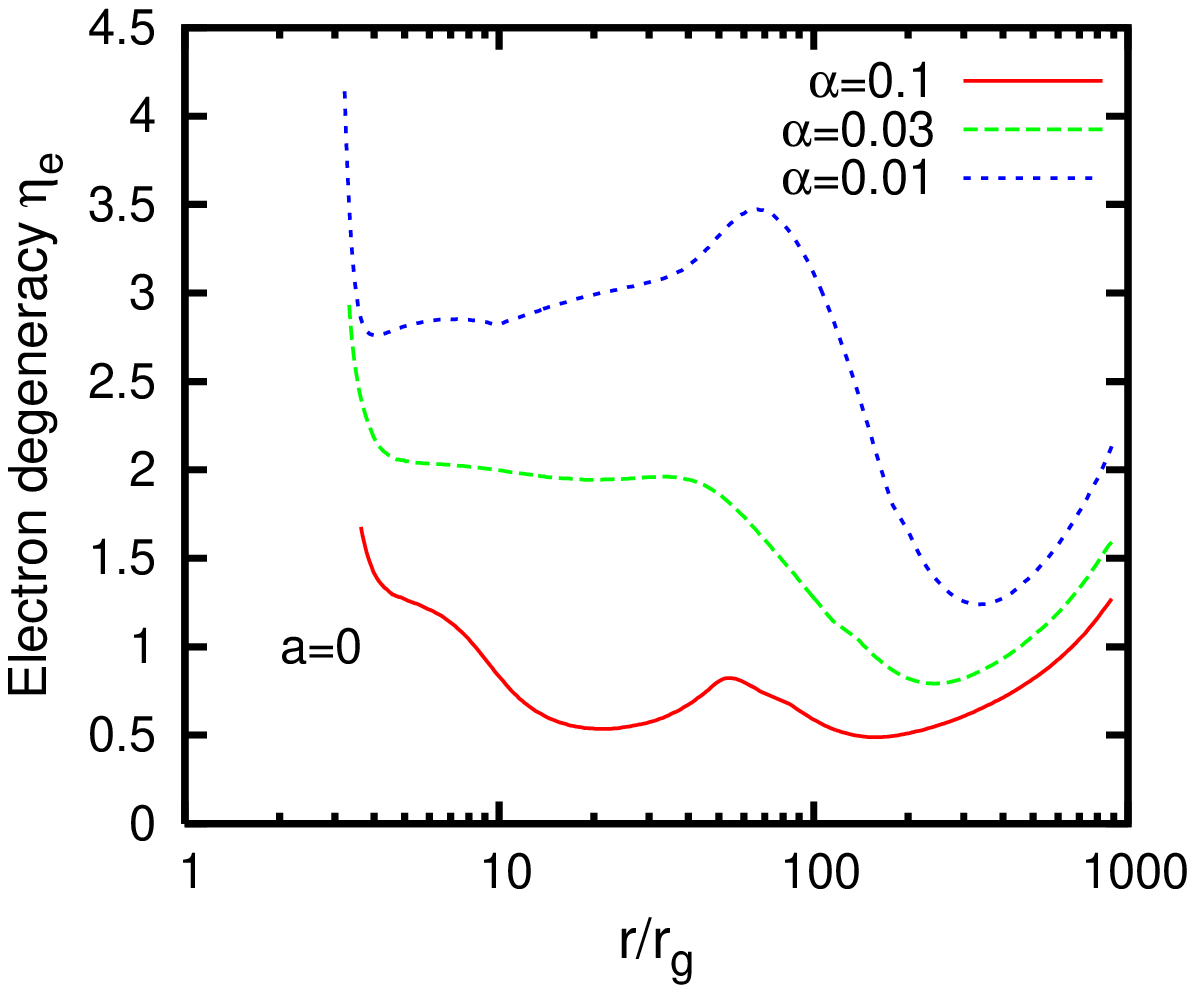}{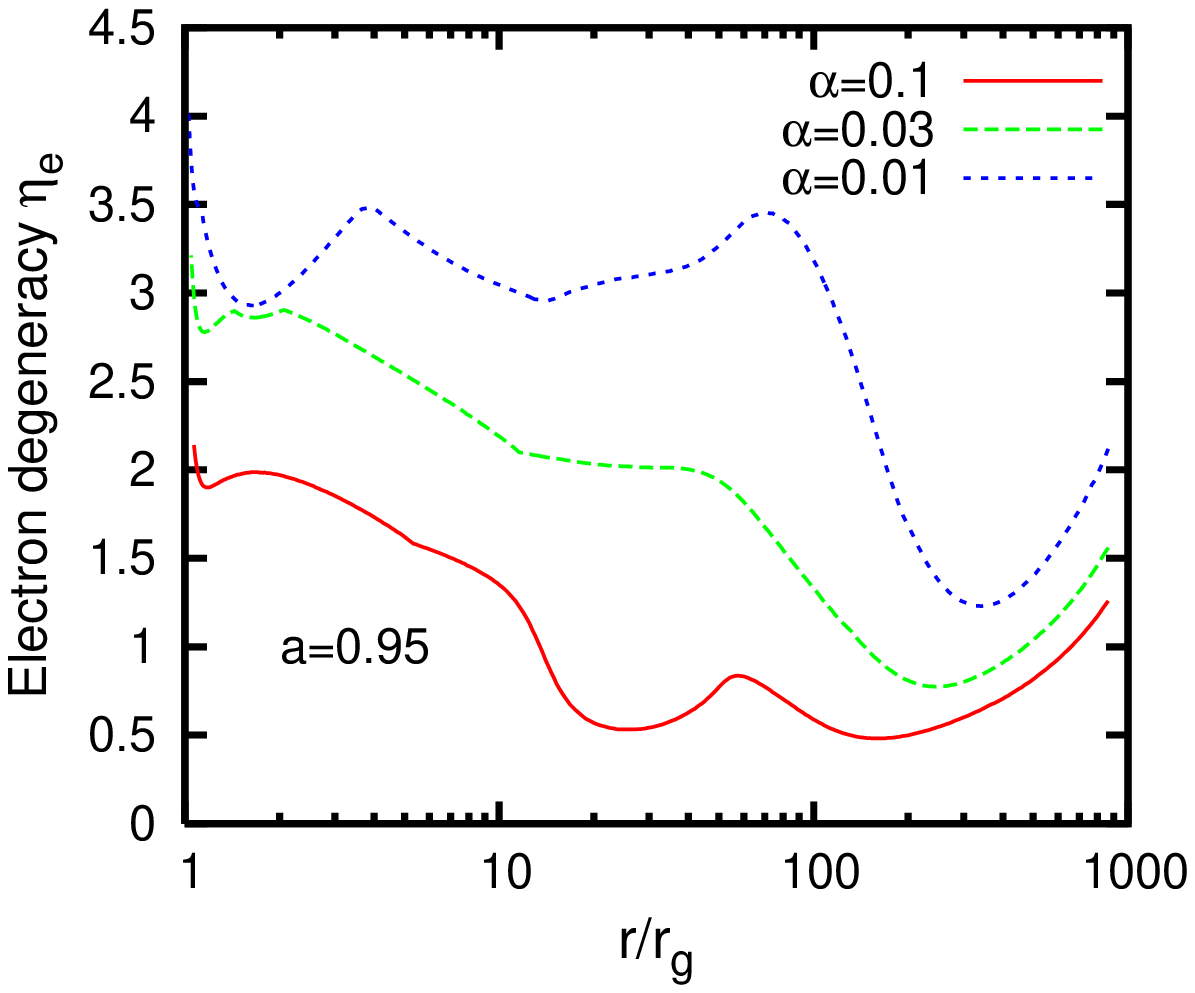}
\label{fig:eta}
\caption{Degeneracy parameter $\eta_e(r)$ for the same disk models as in 
Fig.~1.}
\end{center}
\end{figure}

\begin{figure}
\begin{center}
\plottwo{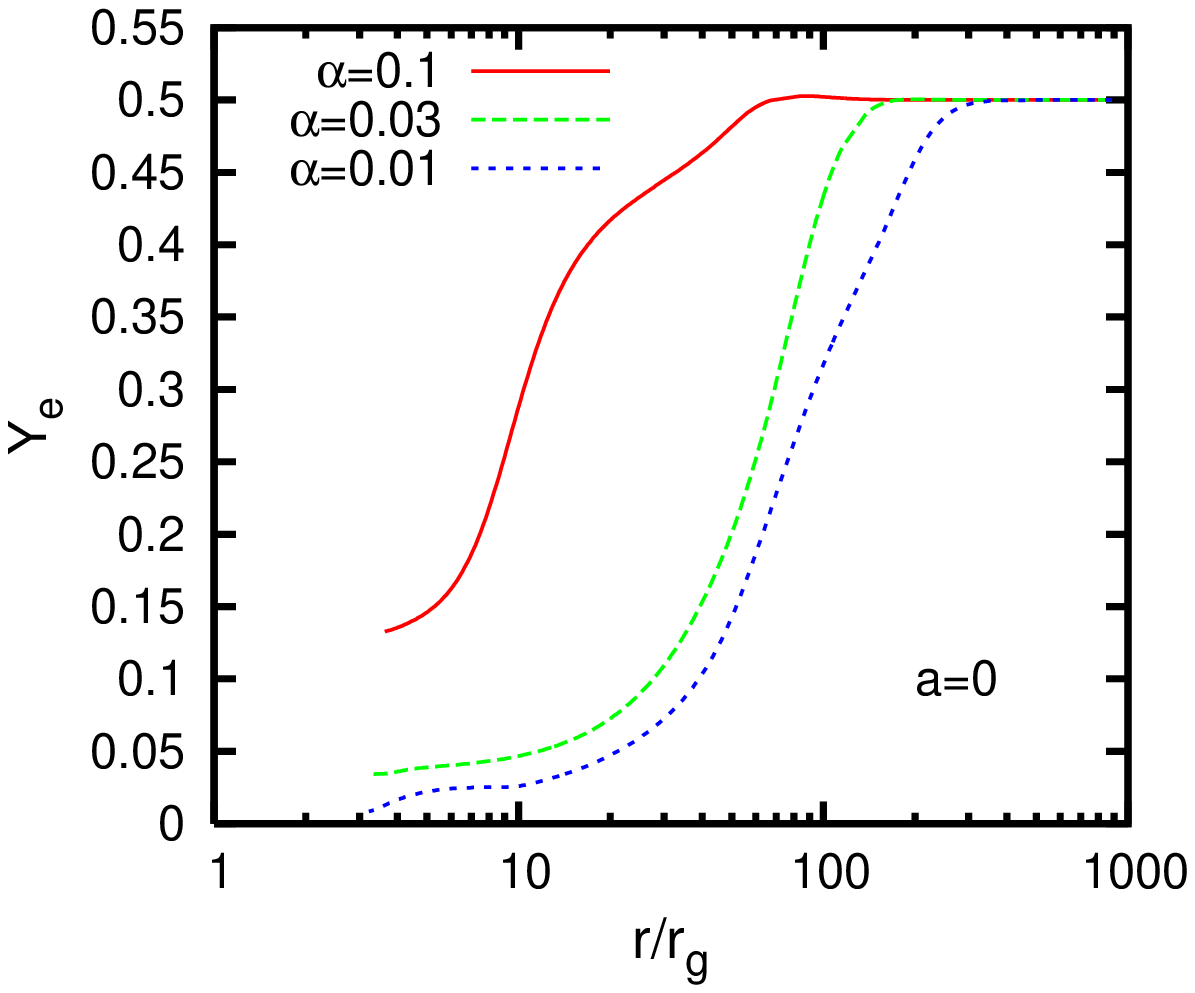}{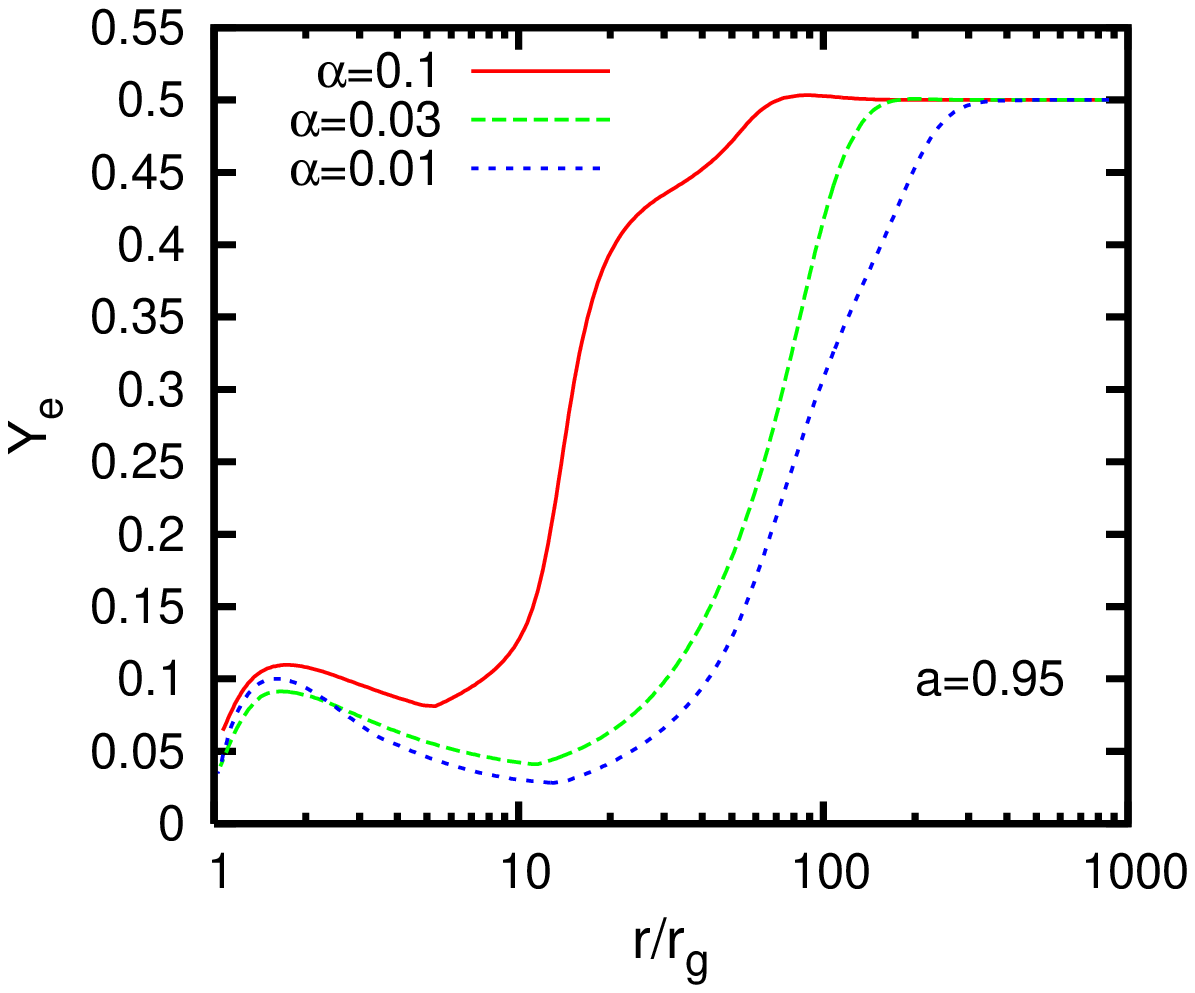}
\caption{$Y_e(r)$ for the same disk models as in Fig.~1.}
\label{fig:Ye}
\end{center}
\end{figure}

\begin{figure}
\begin{center}
\plottwo{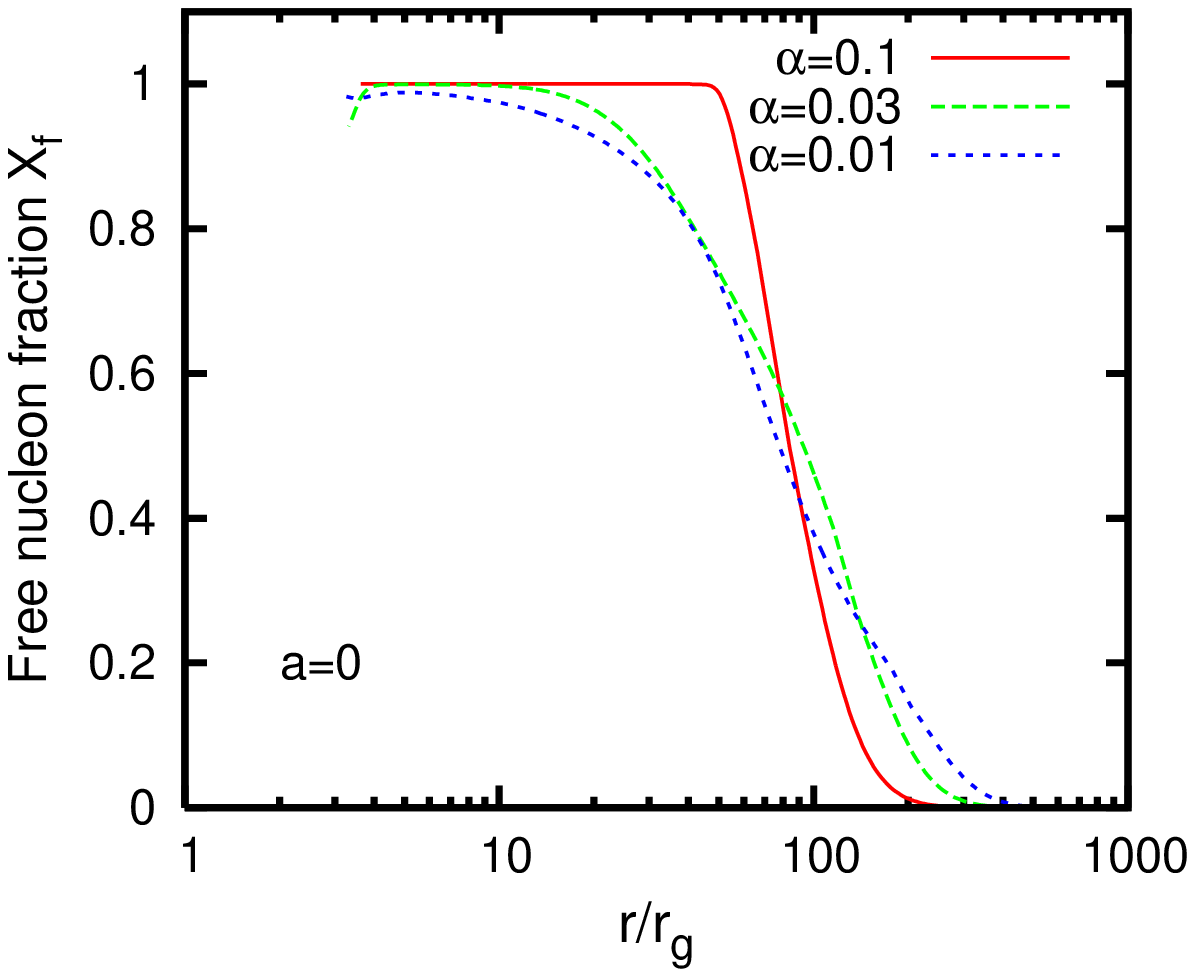}{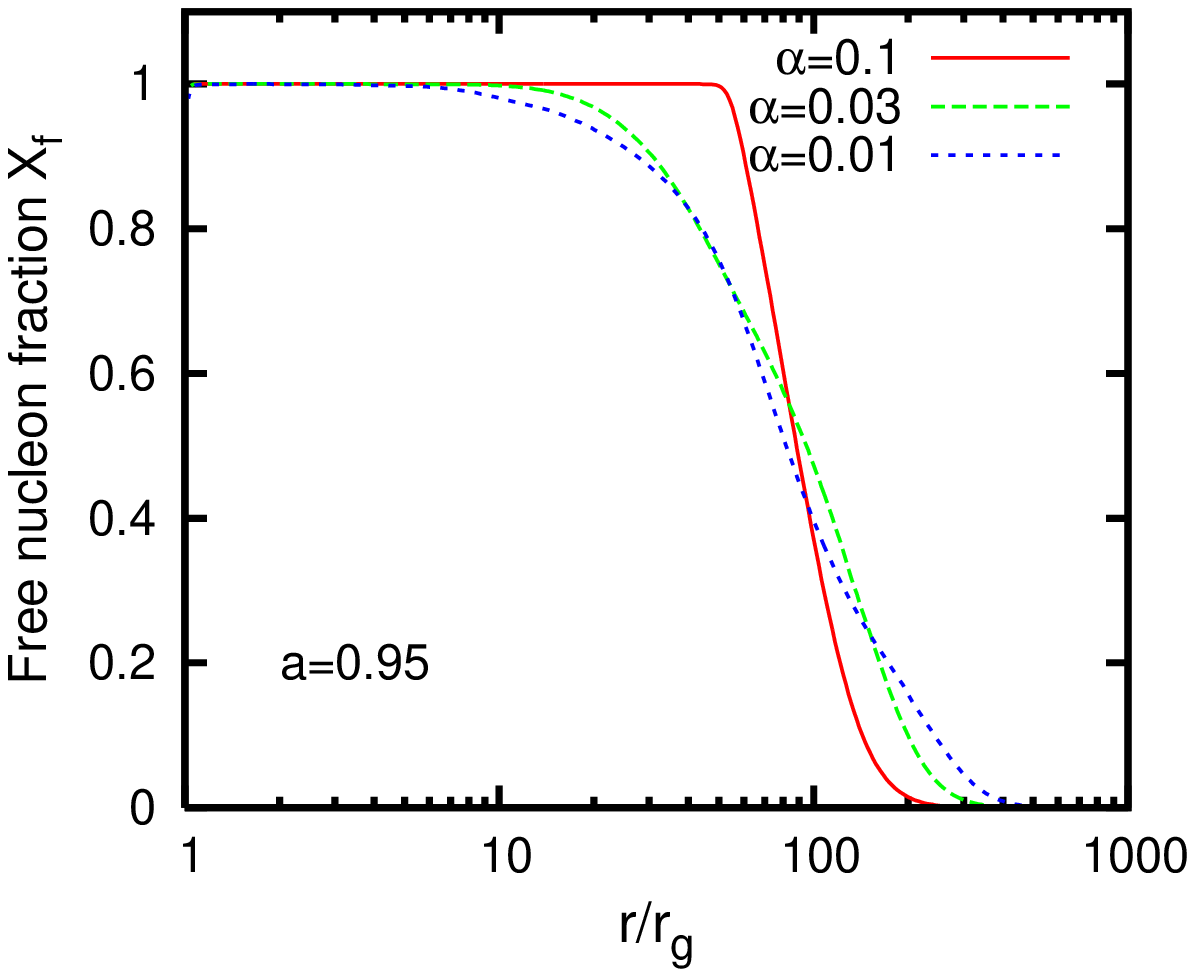}
\label{fig:Xf}
\caption{Mass fraction of free nucleons $X_f(r)$ for the same disk models 
as in Fig.~1.}
\end{center}
\end{figure}

\begin{figure}
\begin{center}
\plottwo{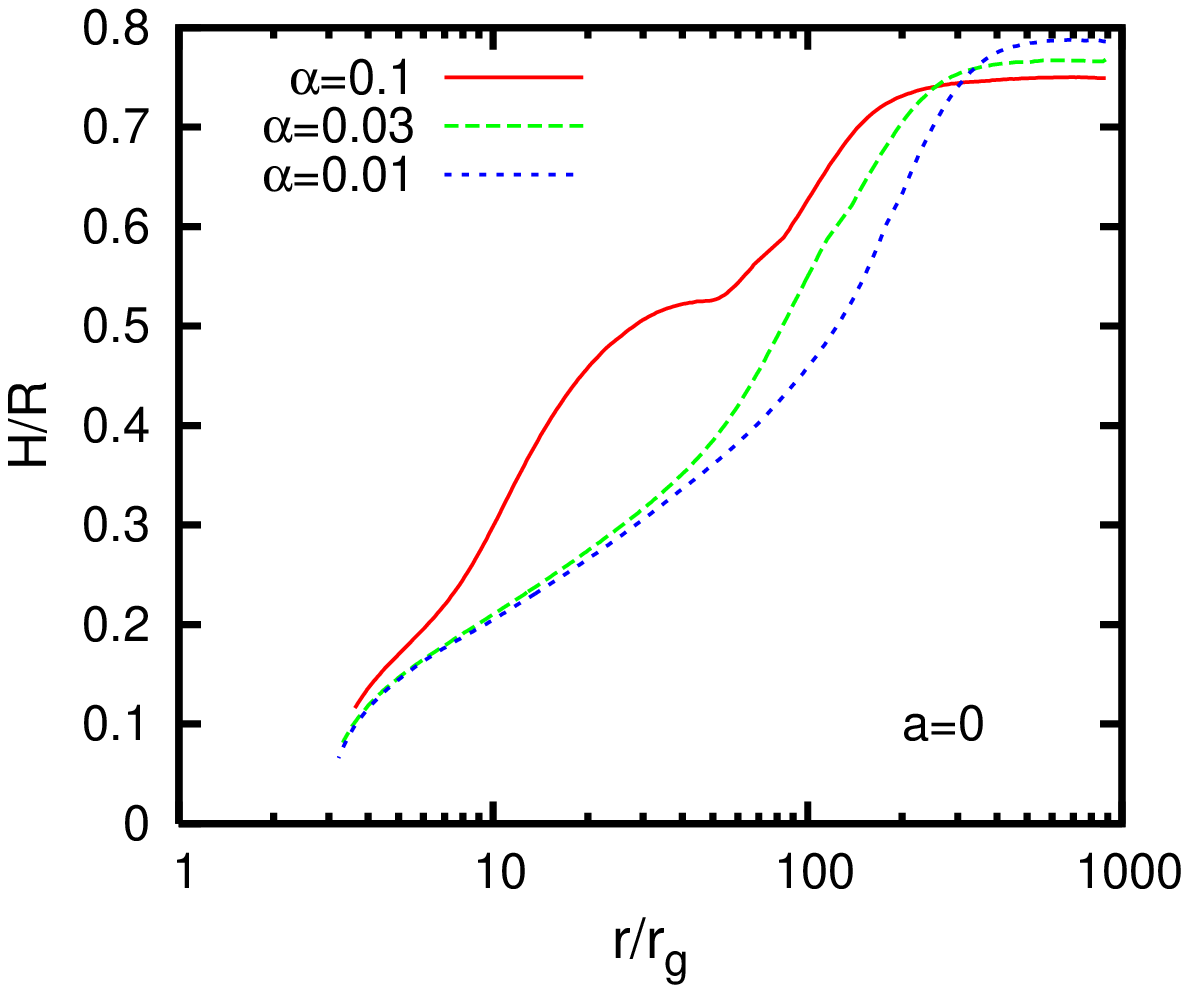}{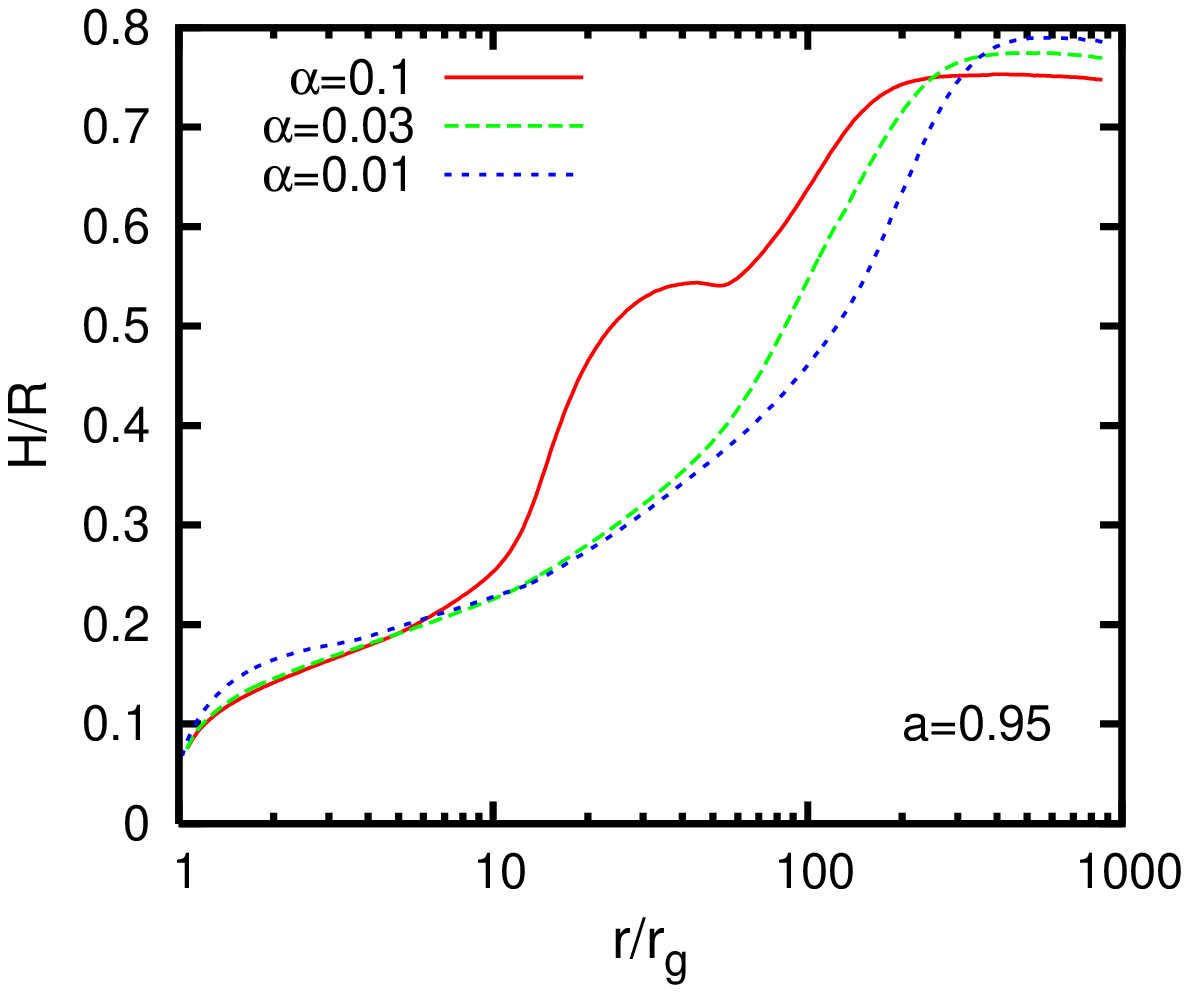}
\label{fig:H}
\caption{Scale-height of the disk $H(r)/r$ for the same disk models as in 
Fig.~1.}
\end{center}
\end{figure}

\begin{figure}
\begin{center}
\plottwo{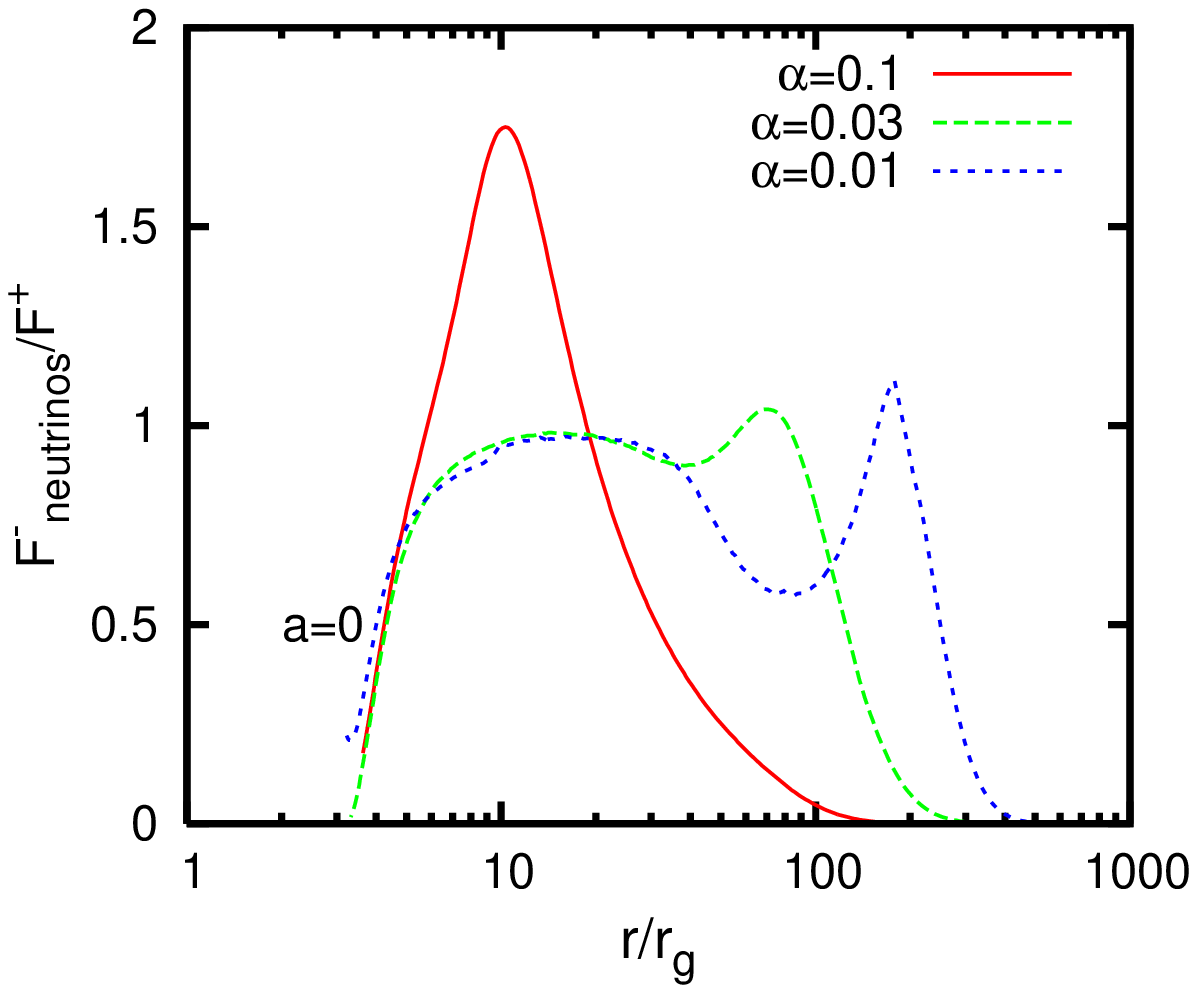}{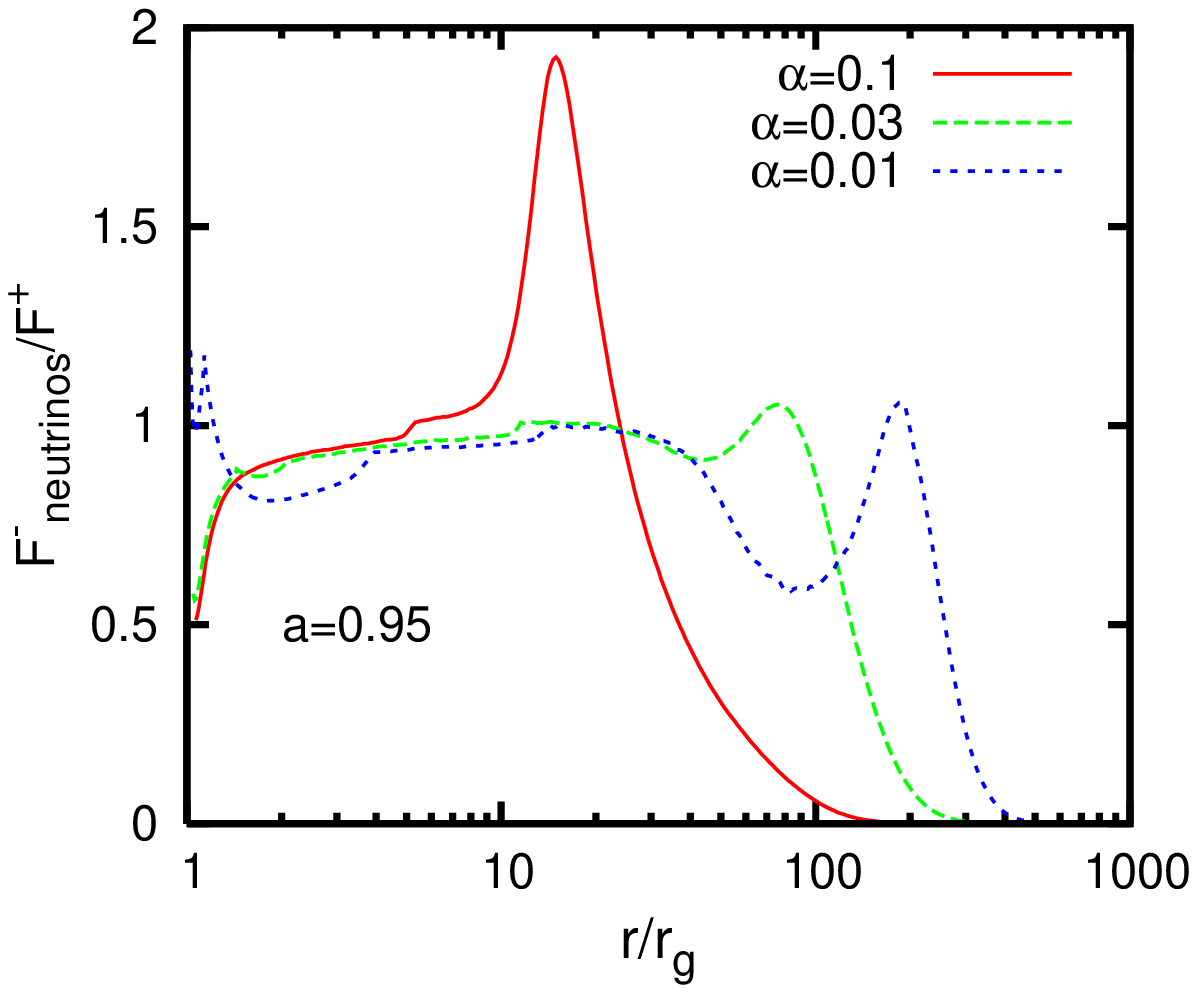}
\label{fig:F}
\caption{Ratio of neutrino flux $F_\nu+F_{\bar\nu}$ to the heating rate
$F^+$ for the same disk models as in Fig.~1.}
\end{center}
\end{figure}

\begin{figure}
\begin{center}
\plottwo{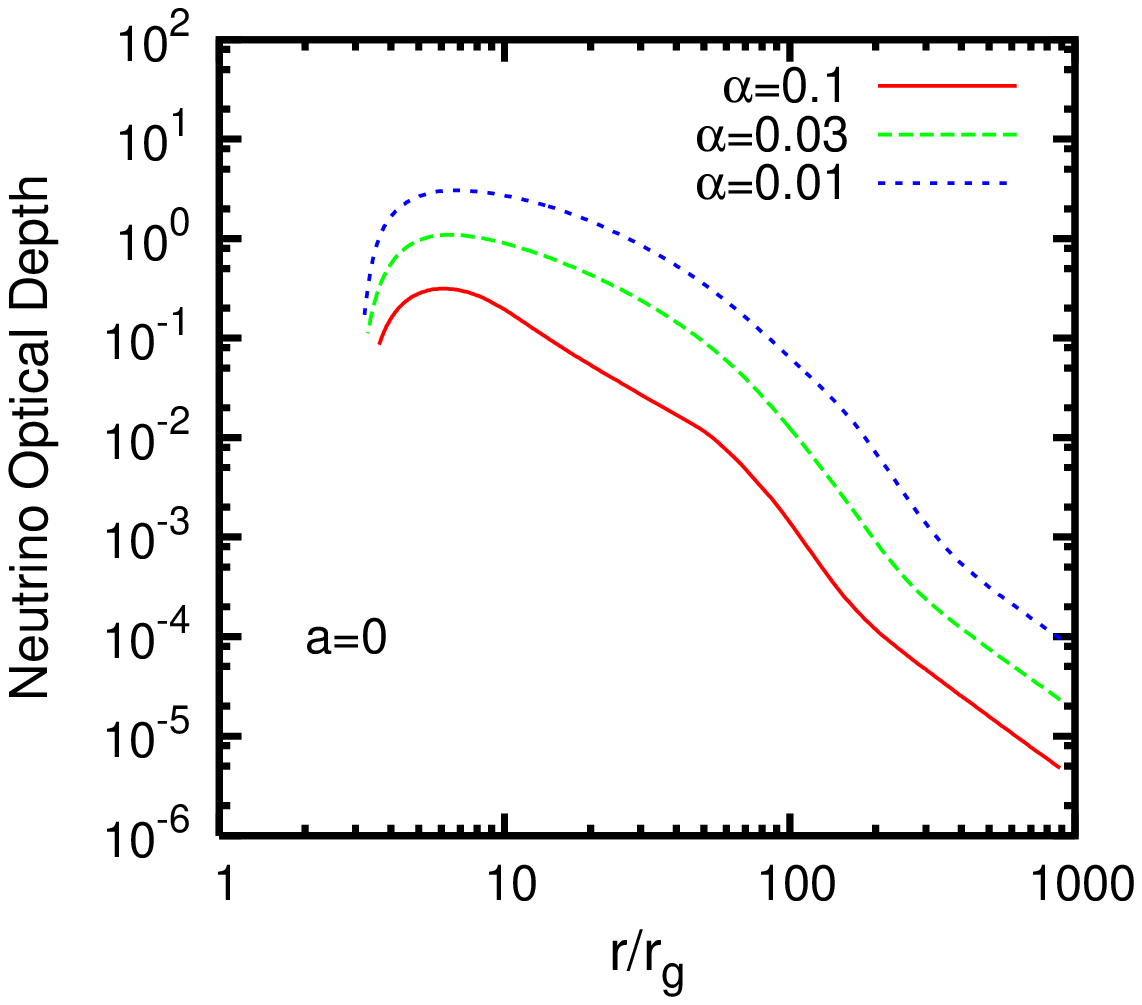}{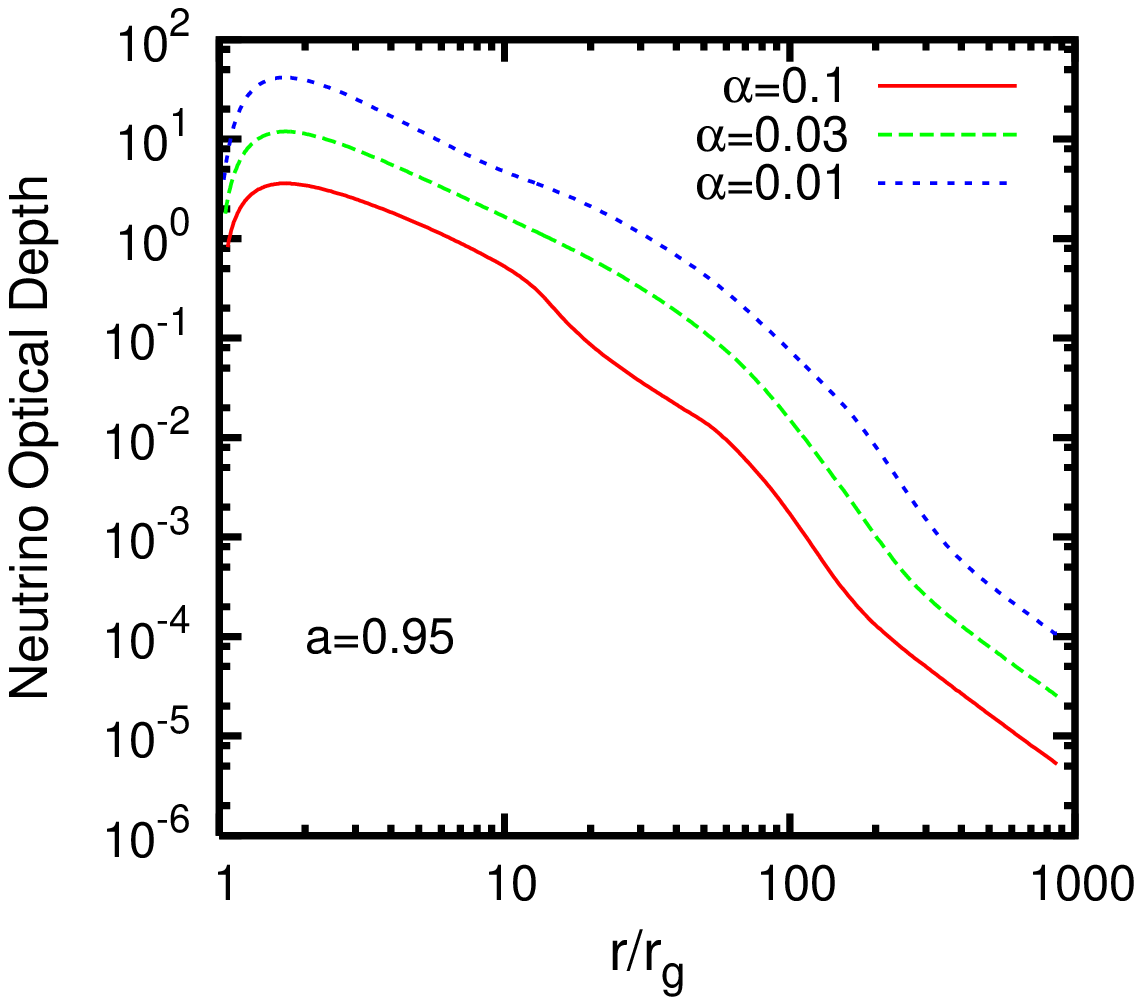}
\label{fig:tau}
\caption{Optical depth seen by neutrinos $\tau_\nu(r)$ for the same disk 
models as in Fig.~1.}
\end{center}
\end{figure}


\begin{figure}
\begin{center}
\plottwo{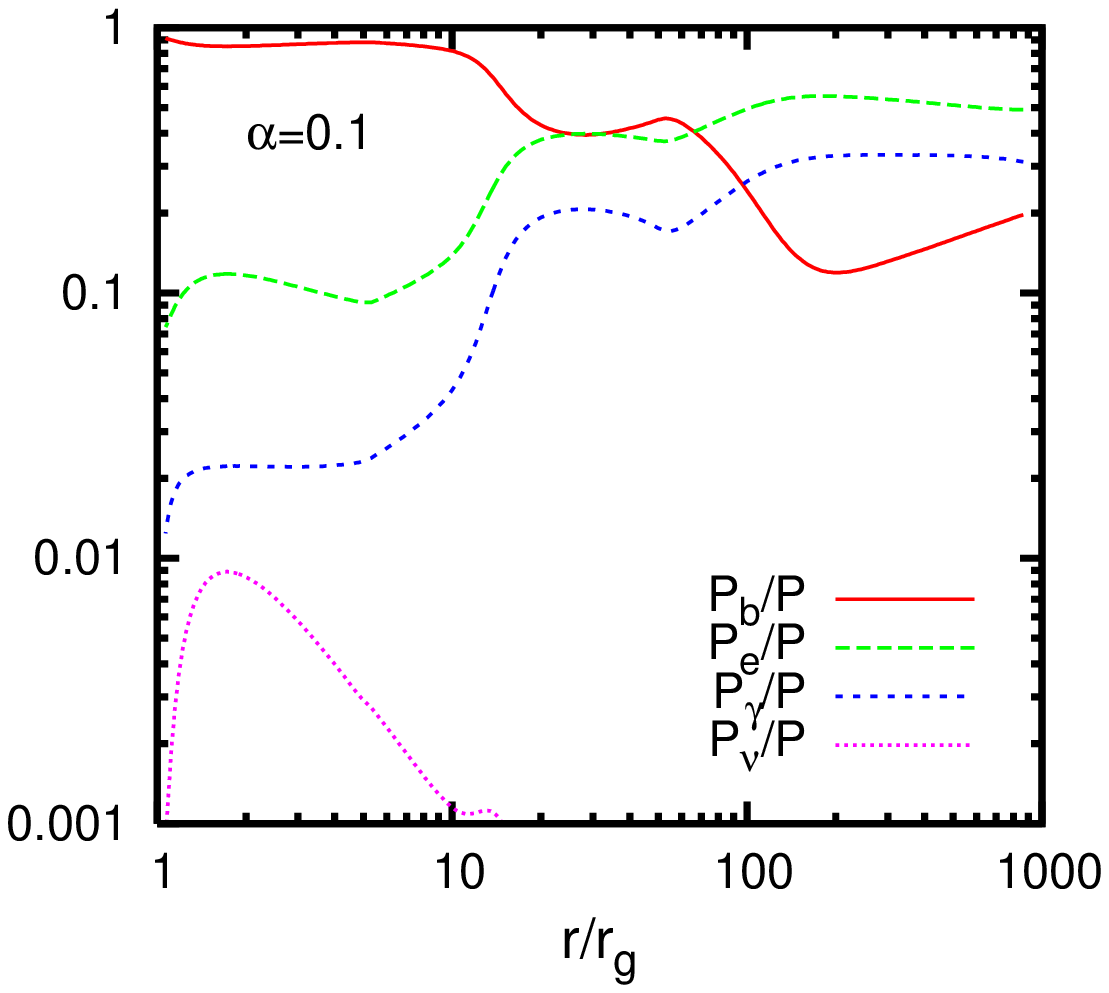}{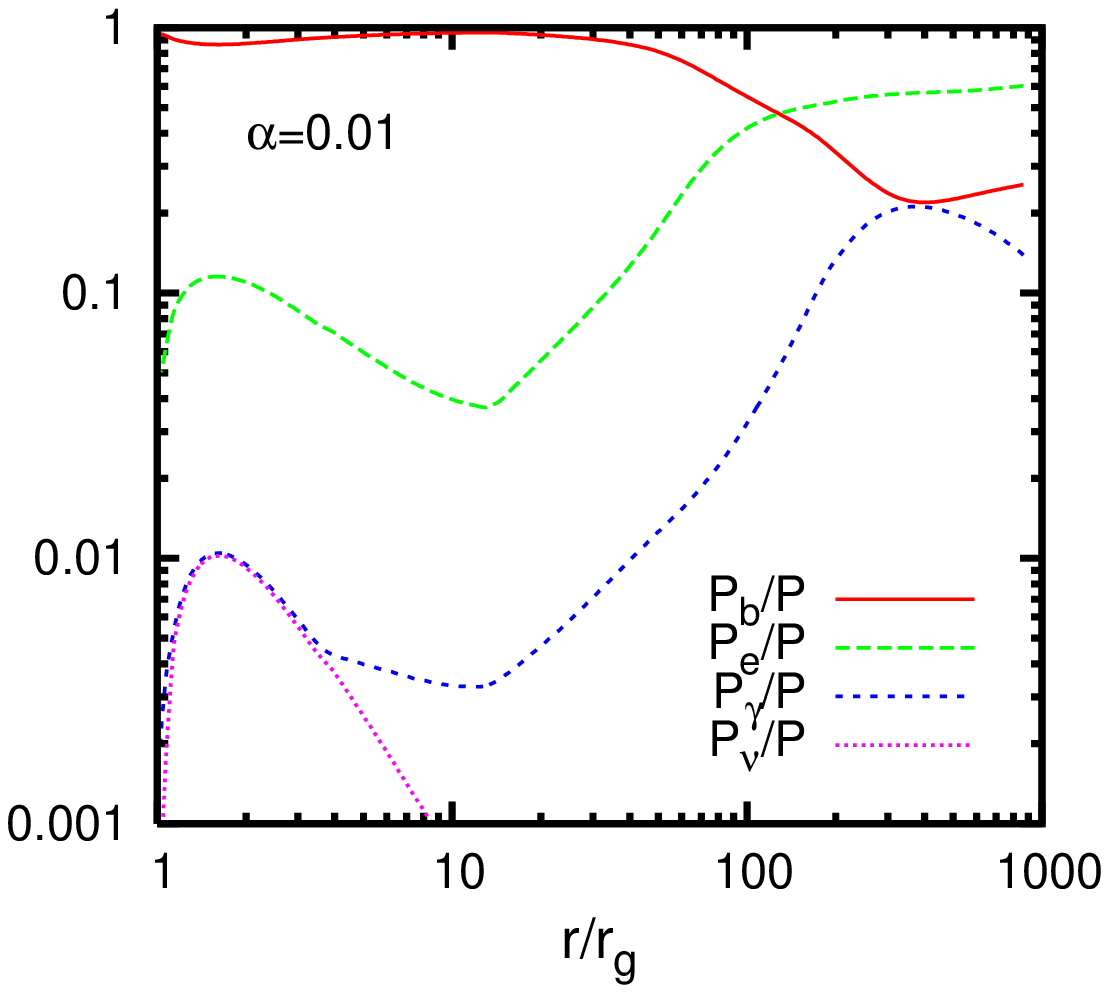}
\label{fig:tau}
\caption{Contributions to total pressure $P$ from baryons,  $P_b$,
electrons and positrons $P_e=P_{e^-}+P_{e^+}$, radiation $P_\gamma$,
and neutrinos $P_\nu+P_{\bar\nu}$ for the Kerr accretion disk ($a=0.95$) 
with $\dM=0.2M_\odot$~s$^{-1}$. {\it Left panel:} model with viscosity
parameter $\alpha=0.1$. {\it Right panel:} model with $\alpha=0.01$. }
\end{center}
\end{figure}


\subsection{Sample Models}
\label{sc:sample}

Figures~1-8 show examples of the disk structure 
for accretion rate $0.2M_\odot$~s$^{-1}$ and three 
different values of viscosity parameter $\alpha=0.1$, $0.03$, and $0.01$. 
For comparison two cases are shown: $a=0$ (Schwarzschild) and 
$a=0.95$ (Kerr). The black hole mass $M=3M_\odot$ is assumed in all models.

First, we note significant differences between models with high 
$\alpha=0.1$ and low $\alpha=0.01$.
Disks with low $\alpha$ accrete slower and have higher density. 
This has the following implications:
(1) the region of neutrino emission in low-$\alpha$ disks 
extends to larger radii $r\sim 200r_g$, 
(2) electron degeneracy $\eta_e=\mu_e/kT$ is higher, and (3) $Y_e$ is lower.

Electron degeneracy $\eta_e$ is an important physical parameter that affects
$Y_e$, pressure, and neutrino cooling.
In the outer advective region, $\eta_e$ is decreasing as the 
heated fluid approaches the black hole until it reaches radii of a few
hundred $r_g$ where nuclear cooling $\Fnuc$ becomes significant.
When only $\sim 10$\% of $\alpha$-particles are disintegrated (see Fig.~5),
an energy $X_f\times 7.1$~MeV$\approx 0.7$~MeV is consumed per nucleon,
which is comparable to the available heat stored in the flow at these
radii. Then $\eta_e$ begins to grow. 

The evolution of $\eta_e$ around $100r_g$ is shaped by the competition of a 
few effects that are sensitive to $\alpha$. In low-$\alpha$ disks, 
neutrino emission becomes significant at $r\sim 200r_g$, which 
implies additional cooling (besides $\Fnuc$) and the drop of $Y_e$ from 
0.5 toward a low equilibrium value. The coupled evolution of $Y_e$, 
$\eta_e$, and neutrino emissivity reaches $\beta$-equilibrium at 
$r\sim 50r_g$. In high-$\alpha$ disks, neutrino emission is less efficient
at $r\sim 100r_g$ and the drop of $Y_e$ occurs at smaller $r$, which
leads to a different evolution of degeneracy $\eta_e$ with radius.
In all cases, however, $\beta$-equilibrium is established with a low $Y_e$
as soon as neutrino emission becomes significant in the energy balance
of the disk.

The nuclear cooling becomes negligible 
where 90\% of $\alpha$-particles disappear (Fig.~5).  
In the $\alpha=0.1$ model, the neutrino cooling is still weak at
this radius and the accreting fluid is quickly heated by $F^+$.
This recovery of disk heating leads to the knee in the profile of 
$H/r$ at $r\approx 40r_g$ (Fig.~6). Then the growth of $T$ above 1~MeV 
ignites strong neutrino emission that carries away the generated heat
and produces the spike of $F_\nu(r)$ between  $10r_g$ and $20r_g$,
overshooting $F^+$ by a factor of 2 (Fig.~7). This overshooting effect 
does not happen in the low-$\alpha$ disks because they are cooled more 
efficiently around $100r_g$.

The Kerr disk models with $a=0.95$ extend to a small radius $\rms\approx r_g$
and have an extended neutrino-cooled region where
$F^+\approx F^-\approx F_\nu+F_{\bar\nu}$, i.e. the approximate
local balance between heating and cooling is established.
This balance requires smaller scale-height $H/r$ at smaller radii. 
As a result, $H/r$ is reduced below 0.2.
Significant differences between Schwarzschild and Kerr cases are observed
in the inner region $r<10r_g$ where most of accretion energy is released. 
The Kerr disk becomes opaque to neutrinos, and the value of $Y_e$ converges
to $Y_e\sim 0.1$ for all three viscosity parameters
$\alpha=0.1$, $0.03$, and $0.01$.

Figure~9 shows the contributions of $P_b$, $P_\gamma$, $P_e=P_{e^-}+P_{e^+}$,
and $P_\nu$ to the total pressure $P$ for two models of Kerr disk. 
One can see that the baryon pressure $P_b$ dominates in the 
neutrino-cooled region. This is a general property of all neutrino-cooled 
disks, which corresponds to their
mild degeneracy $\eta_e\sim 1-3$. In the limit of strong electron 
degeneracy, $P_e$ would be dominant. In the limit of low degeneracy, 
$P_\gamma\sim P_e$ would be dominant. Only at mild degeneracy
$P_\gamma$ and $P_e$ are small compared with $P_b$. The dominance 
of $P_b$ is thus a special feature of $\nu$-cooled disks that is related 
to their self-regulation toward the mild degeneracy and low $Y_e\sim 0.1$.

\subsection{Survey of Models}

A schematic picture of accretion disk is shown in Figure~10. 
The disk has 5 characteristic radii:

1. --- Radius $r_\alpha$ where 50\% of $\alpha$-particles are decomposed
into free nucleons. The destruction of $\alpha$-particles consumes 7~MeV 
per nucleon, which makes the disk thinner. 

2. --- "Ignition" radius $\rign$ where neutrino emission switches on. 
At this point, the mean electron energy becomes comparable to 
$(m_n-m_p)c^2$, enabling the capture reaction $e^-+p\rightarrow n+\nu$. 
Then neutrino cooling becomes 
significant, further reducing the disk thickness $H/r$. We choose
the condition $F_\nu+F_{\bar\nu}=F^+/2$ as a formal definition of $\rign$.

3. --- Radius $r_\nu$ where the disk becomes opaque for neutrinos
and they relax to a thermal distribution. Note that the disk is still 
almost transparent for anti-neutrinos at this radius.

4. --- Radius $r_{\bar\nu}$ where the disk becomes opaque for anti-neutrinos,
so that both $\nu$ and ${\bar\nu}$ are now in thermal equilibrium with 
the matter. The disk is still cooled efficiently at this
radius since $\nu$ and ${\bar\nu}$ diffuse and escape the flow faster 
than it accretes into the black hole.

5. --- Radius $\rtr$ where the timescale of neutrino diffusion out of
the disk, $t_{\rm diff}=(H/c)\tau_\nu$ becomes longer than the accretion 
timescale, and neutrinos get trapped and advected into the black hole. 
The transition radius $\rtr$ is formally defined where 
$F_\nu+F_{\bar\nu}$ drops below $F^+/2$.

In addition, there is a radius beyond which the steady-disk model is 
inconsistent because of gravitational instability (e.g. Paczy{\'n}ski 1978). 
We estimate this boundary from the condition
\be
   Q=\frac{c_s\Omega}{2\pi GH\rho}\approx 1.
\ee
The unstable region is where $Q<1$. Besides the gravitational instability, 
the disk mass in this region becomes comparable to that of the black hole, 
and the gravitational potential is not described by the Kerr metric.

\begin{figure}
\begin{center}
\plotone{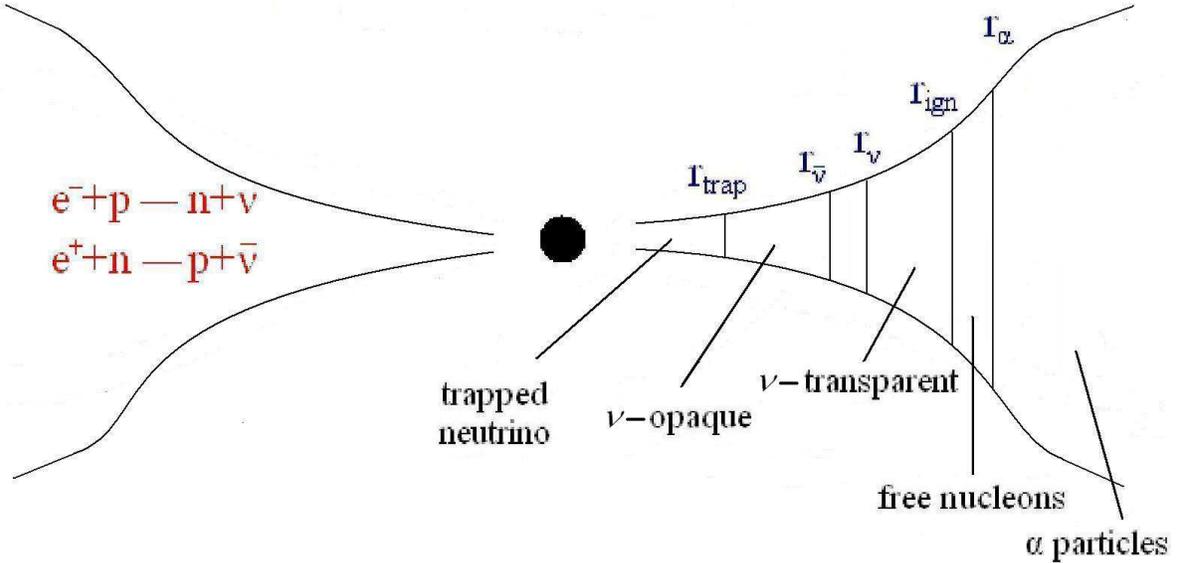}
\caption{Schematic picture of the disk with characteristic radii indicated.}
\end{center}
\end{figure}

We ran a series of models with various $\dM$, for two values of 
the spin parameter $a=0$ and $a=0.95$. For each model, we found the five 
characteristic radii and the region of gravitational instability.
The results are summarized in Figures~11 and 12.
Contour plots of temperature $T$, density $\rho$, electron fraction 
$Y_e$, and efficiency of local cooling $(F_\nu+F_{\bar\nu})/F^+$ 
are shown in Figures~13-16.  

As one can see from Figures~11 and 12, the radius of 50\% disintegration 
of $\alpha$-particles exists at all $\dM$ of the sequence 
($\dM\simgt 10^{-3}M_\odot$~s$^{-1}$) and weakly depends on $\dM$. 
In most models it is between 40 and $100r_g$.

The ignition radius $\rign$ exists if $\dot{M}>\dMign$ which depends on 
$\alpha$ and $a$. Disks with $\dM<\dMign$ remain advective all the way 
to the black hole. For example, for the Kerr disk with $a=0.95$ and 
$\alpha=0.1$, $\dMign\approx 0.02M_\odot$~s$^{-1}$.
By contrast, for the Schwarzschild disk with the same $\alpha=0.1$, 
$\dMign\approx 0.07M_\odot$~s$^{-1}$. 
One can see in Figures~11 and 12 that the ignition radius first appears in 
the inner region when $\dM=\dMign$. As $\dM$ increases 
$\rign$ moves to $\sim 100r_g$. 

The radii of transparency, $r_\nu$ and $r_{\bar\nu}$, scale with $\dM$ 
approximately as $\dM^{3/2}$, and $r_\nu\approx 2r_{\bar\nu}$. 
Transparency of the disk for neutrinos also depends on the black-hole spin. 
The disk becomes opaque if $\dM>\dMop\approx 0.1M_\odot$~s$^{-1}$ 
in the Kerr case ($a=0.95$) and if $\dM>\dMop\approx 1M_\odot$~s$^{-1}$ in 
the Schwarzschild case. 

Significant trapping of neutrinos occurs in the inner region of the Kerr
disk if $\dot{M}>\dMtr$ which also depends on $\alpha$ and $a$.
For example, for $\alpha=0.1$, $\dMtr\approx 2M_\odot$~s$^{-1}$ 
if $a=0.95$ and $\dMtr\approx 10M_\odot$~s$^{-1}$ if $a=0$.
The trapping radius $\rtr$ grows linearly with $\dM$.  

\begin{figure}
\begin{center}
\plottwo{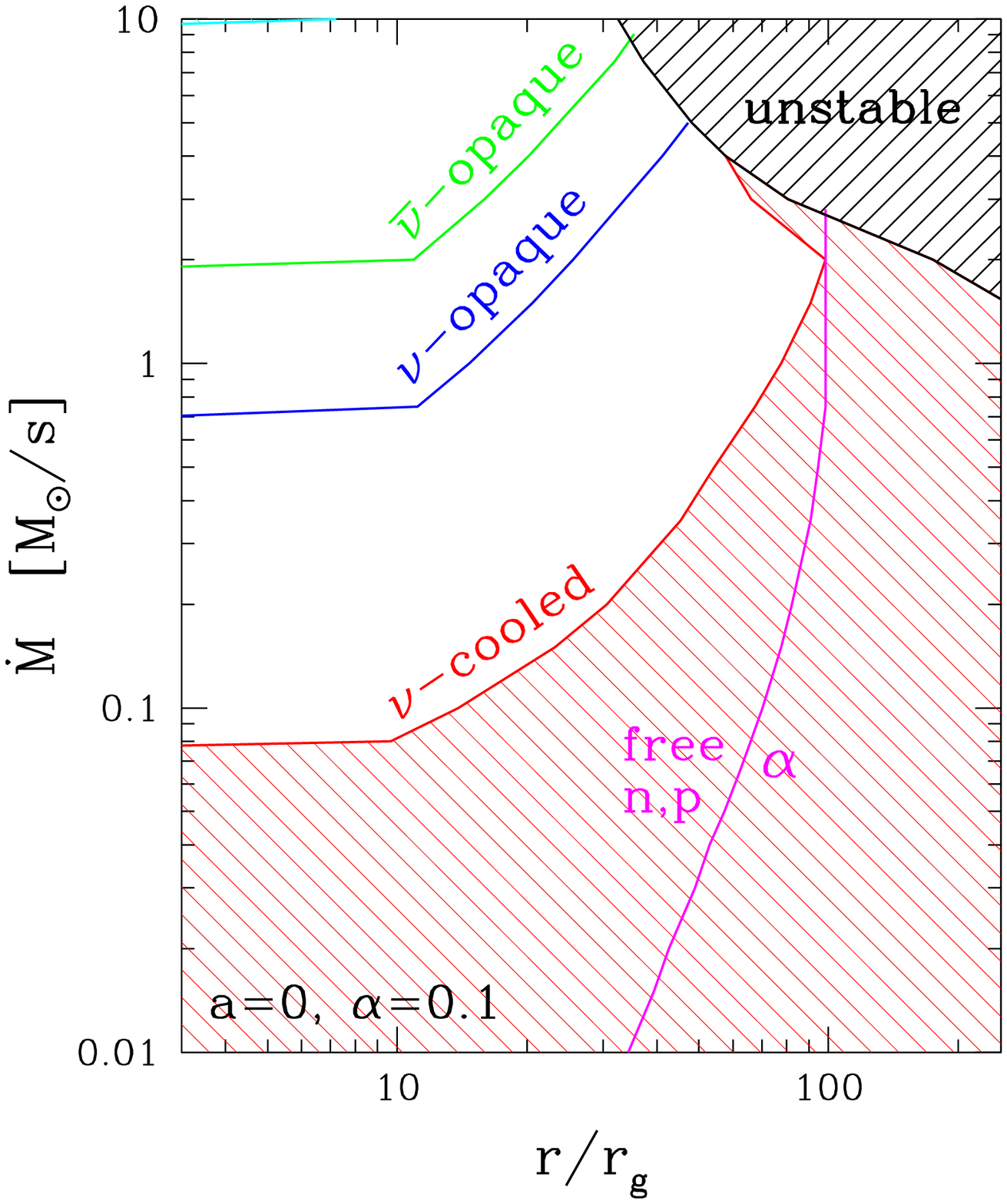}{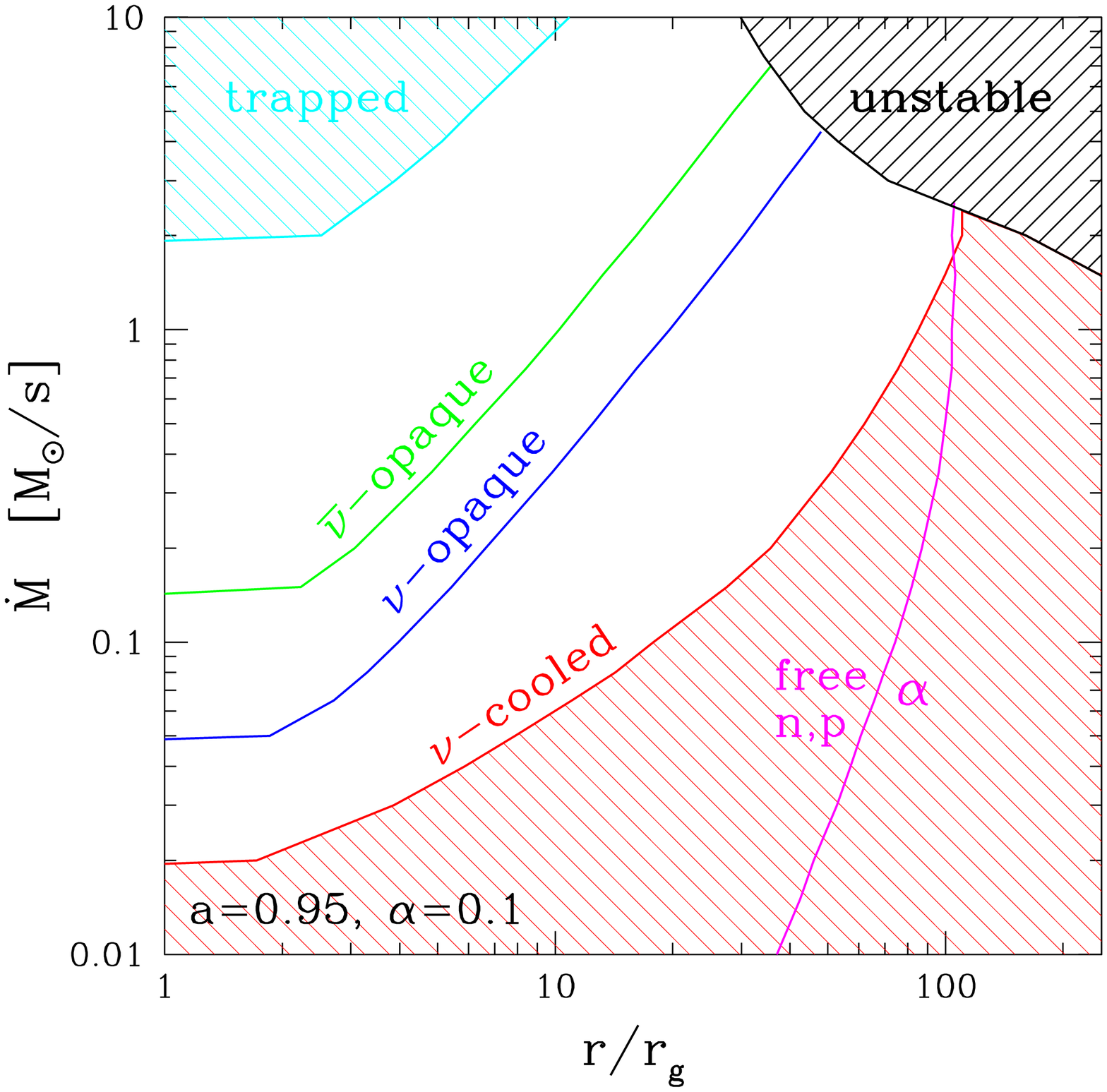}
\caption{
Boundaries of different regions on the $r$-$\dM$ plane for disks around
a black hole of mass $M=3M_\odot$ and spin parameter $a=0$ (left)
and $0.95$ (right). Viscosity parameter $\alpha=0.1$ is assumed.
Neutrino cooling is small in the shadowed region below the "$\nu$-cooled" 
curve and above the ``trapped'' curve. 
The shadowed region marked ``unstable'' is excluded: the steady model is 
inconsistent in this region because of the gravitational instability.
The disk extends down to the marginally stable orbit of radius $\rms=3r_g$
for $a=0$ and $\rms\approx r_g$ for $a=0.95$, where $r_g=2GM/c^2$.
} 
\end{center}
\end{figure} 
\begin{figure}
\begin{center}
\plottwo{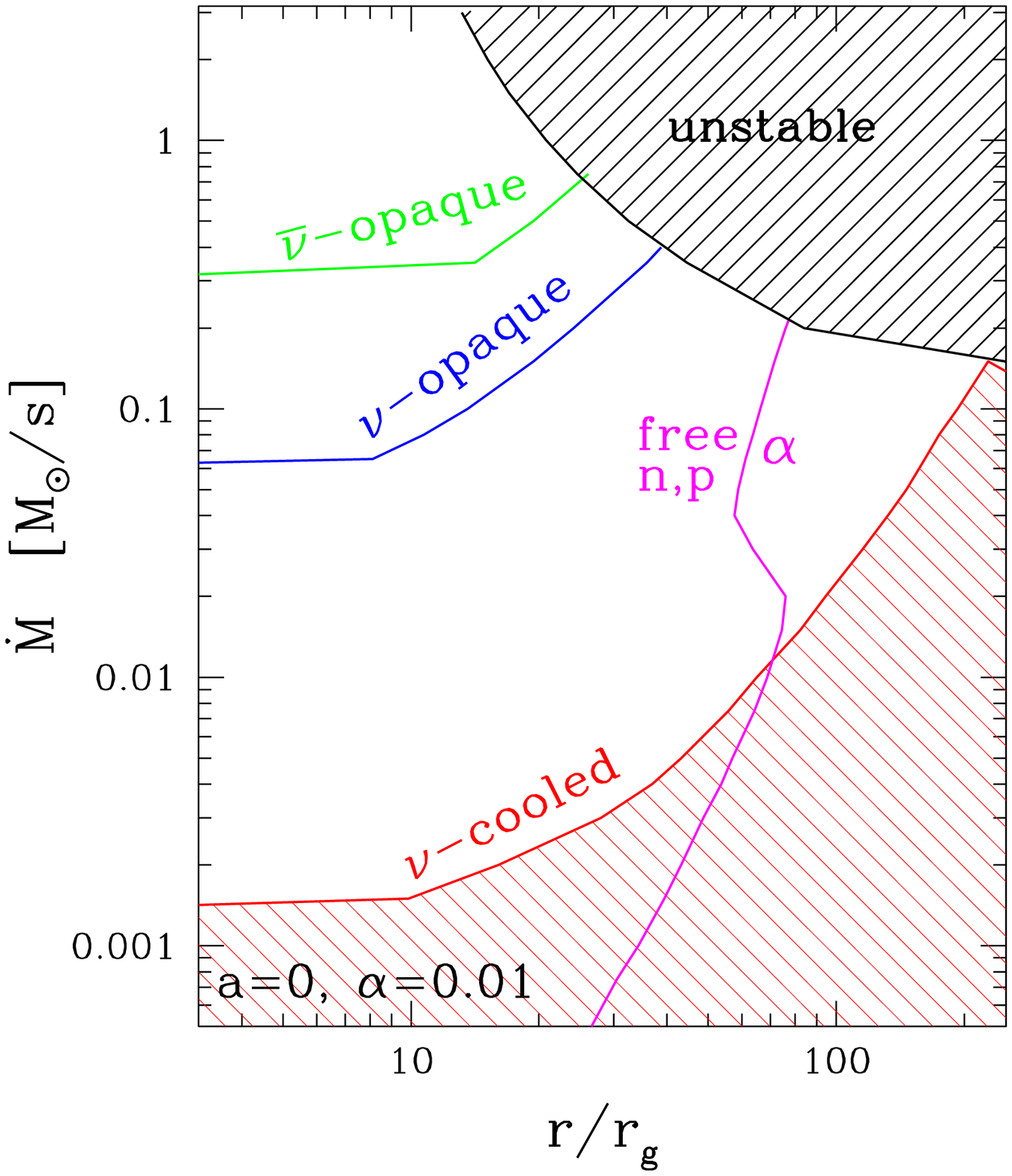}{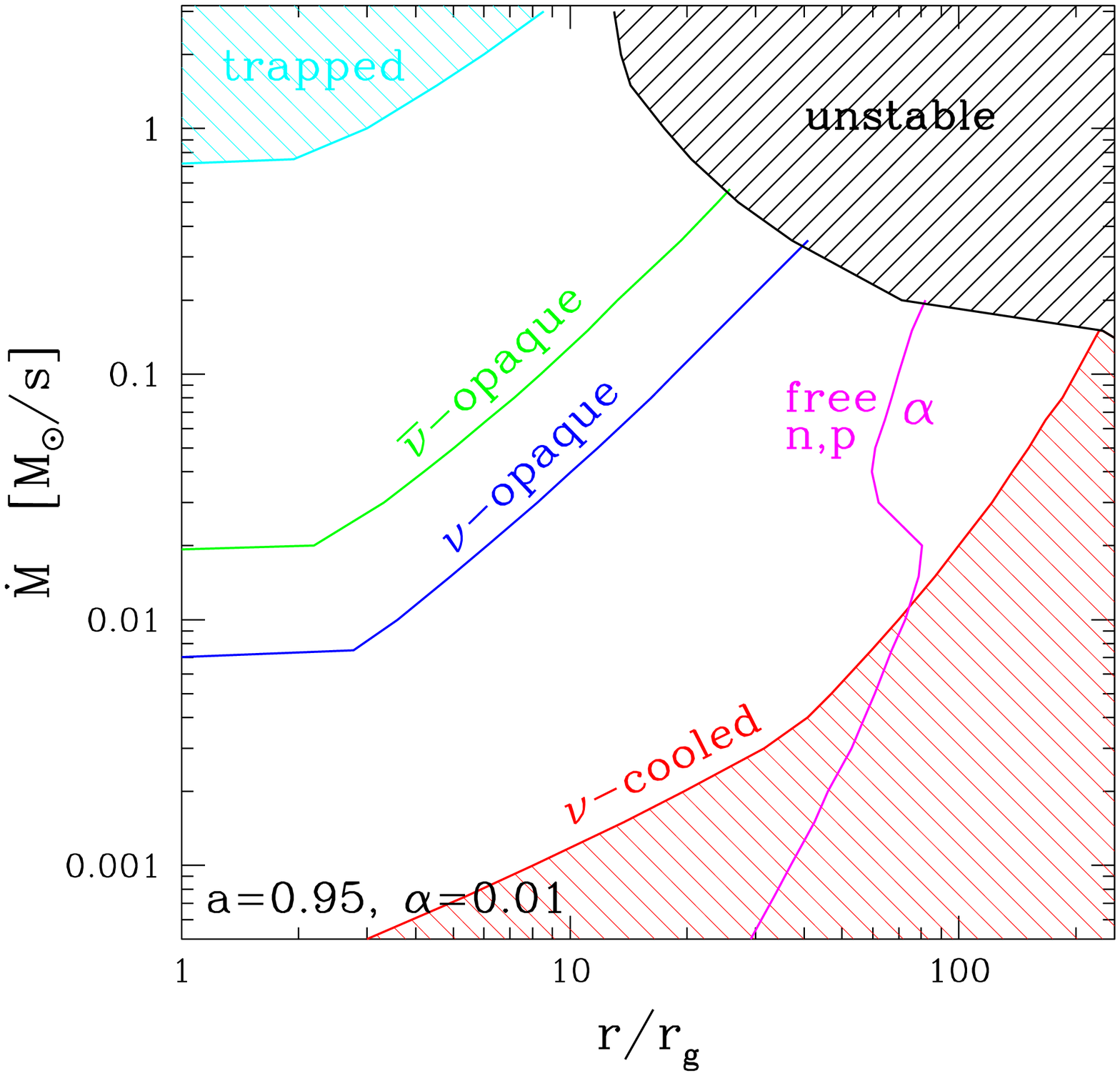}
\caption{
Same as Fig.~11 but for $\alpha=0.01$.
} 
\end{center}
\end{figure} 

\begin{figure}
\begin{center}
\plotone{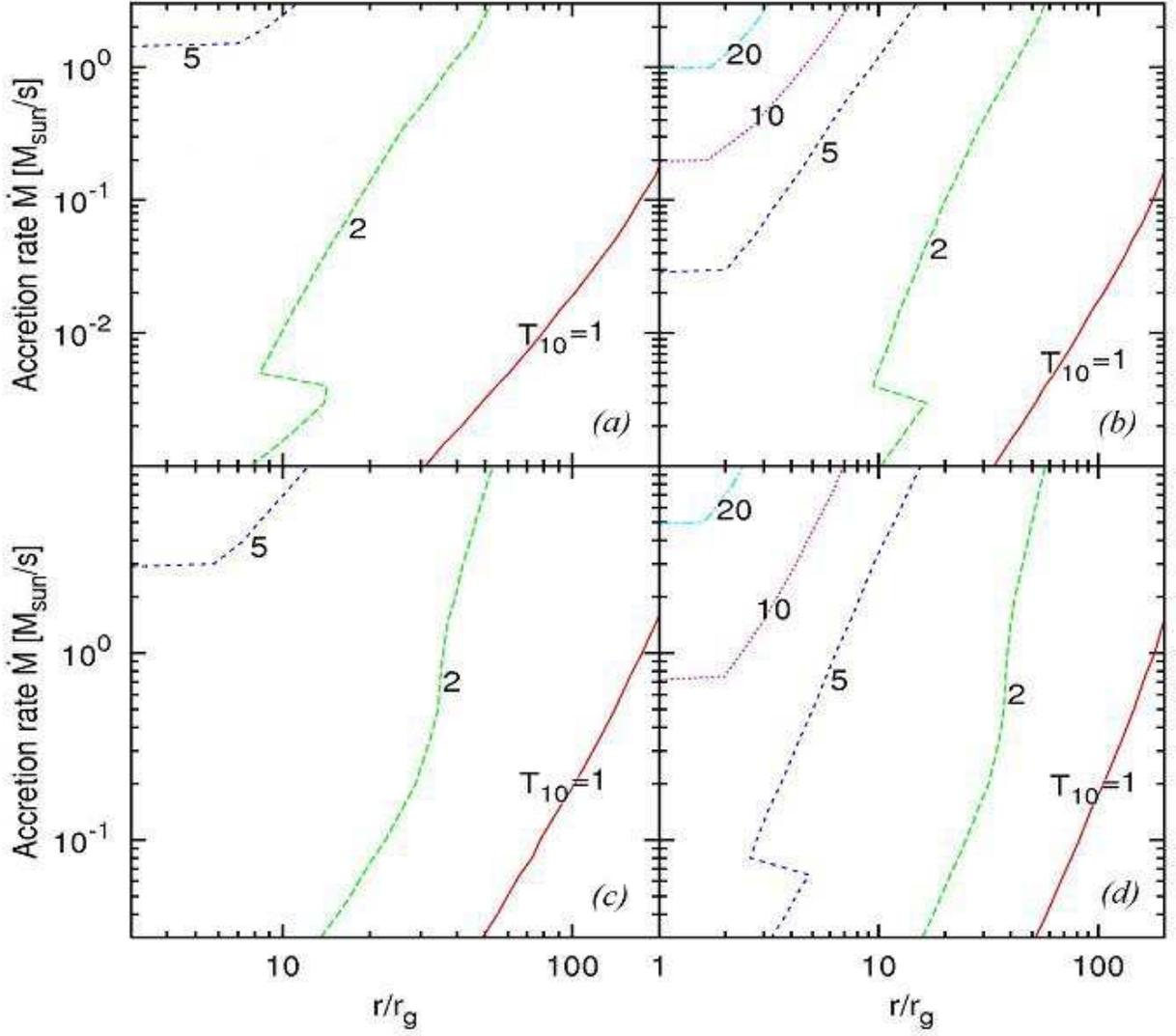}
\caption{
Contours of temperature $T$ (in units of $10^{10}$~K) on the $r-\dM$ plane. 
(a) $\alpha=0.01$ and $a=0$. (b) $\alpha=0.01$ and $a=0.95$. 
(c) $\alpha=0.1$ and $a=0$.  (d) $\alpha=0.1$ and $a=0.95$.
} 
\end{center}
\end{figure} 

\begin{figure}
\begin{center}
\plotone{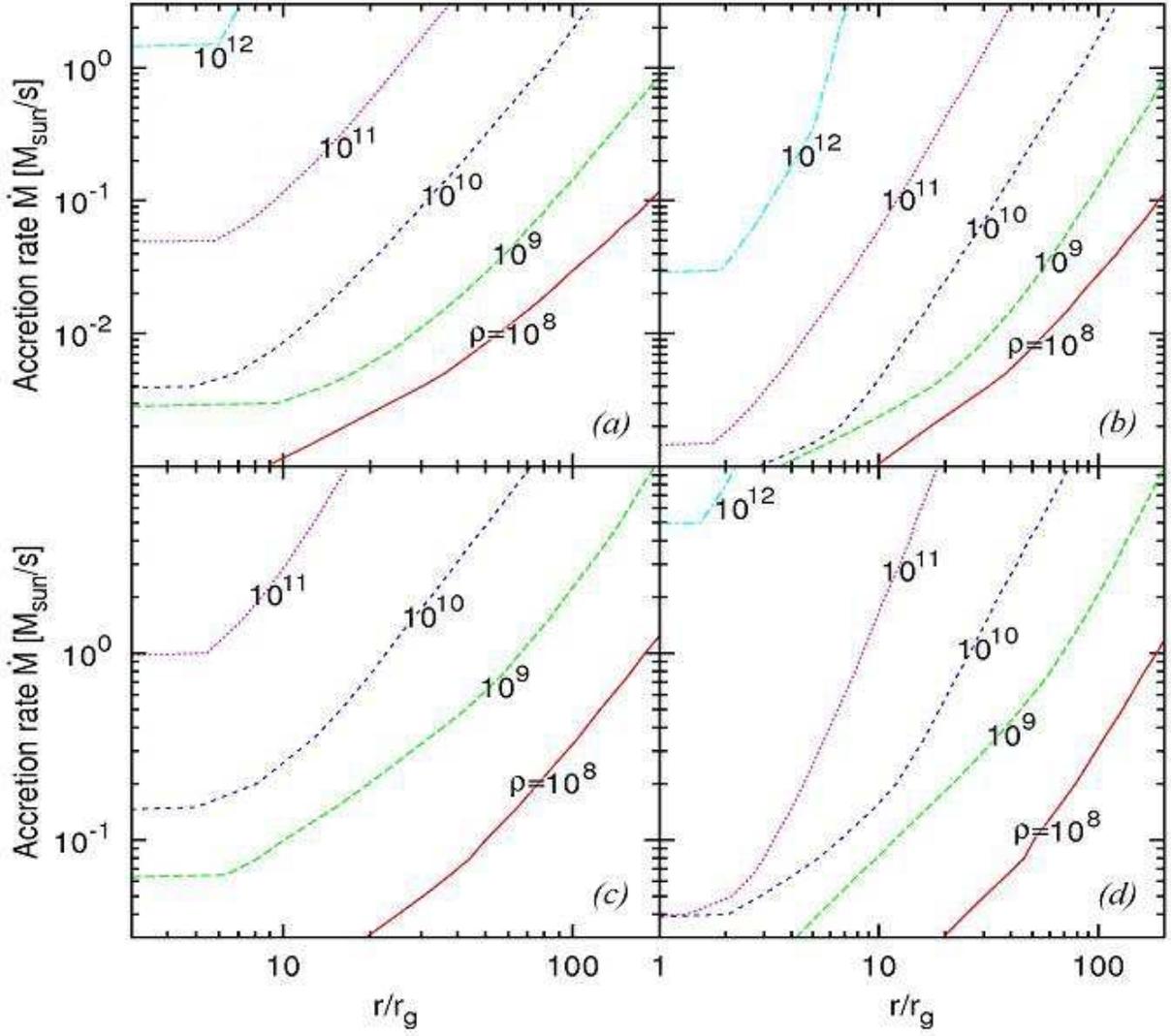}
\caption{
Contours of density $\rho$ (in units of g~cm$^{-3}$) on the $r-\dM$ plane. 
(a) $\alpha=0.01$ and $a=0$. (b) $\alpha=0.01$ and $a=0.95$. 
(c) $\alpha=0.1$ and $a=0$.  (d) $\alpha=0.1$ and $a=0.95$.
} 
\end{center}
\end{figure} 

\begin{figure}
\begin{center}
\plotone{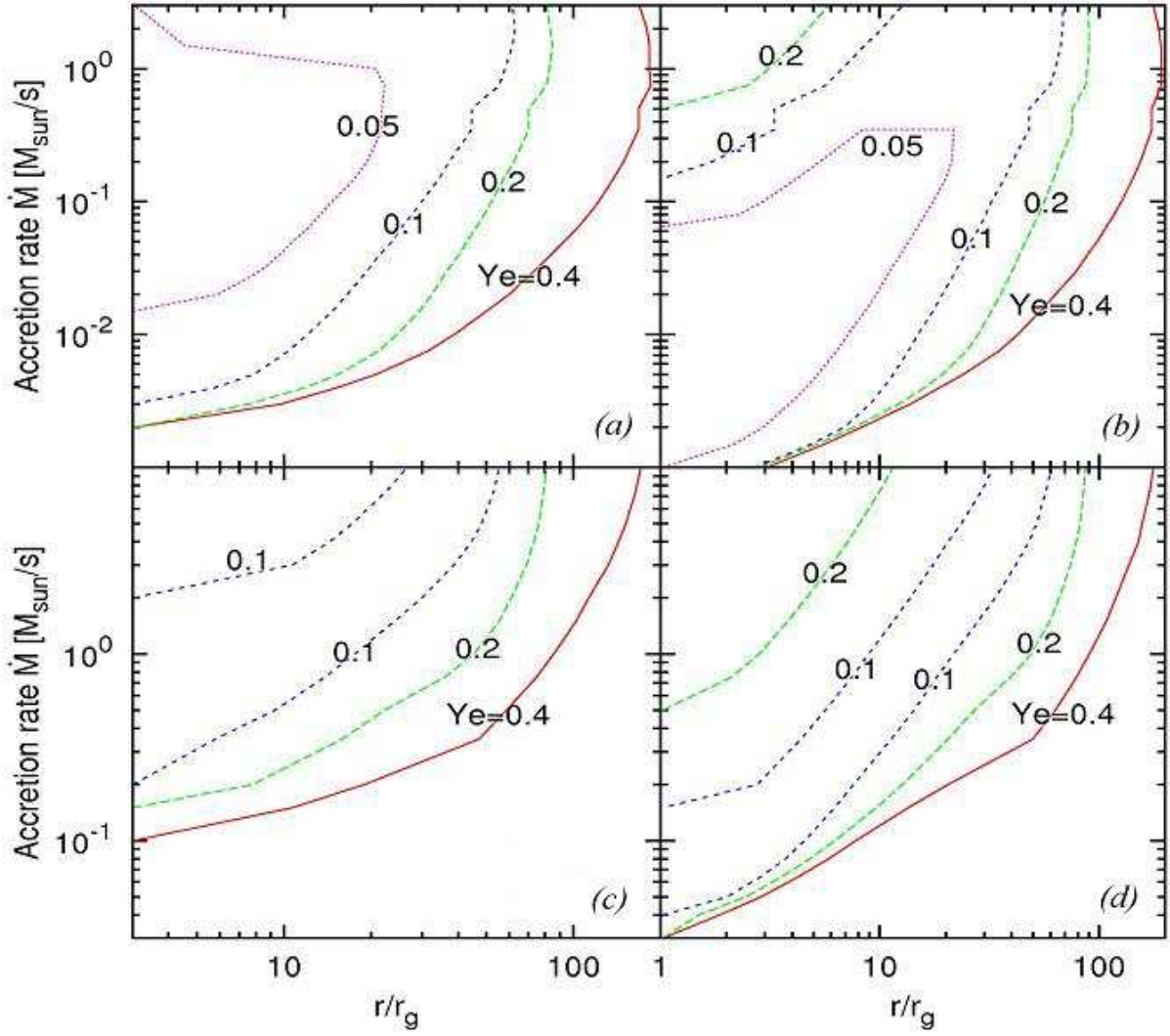}
\caption{
Contours of $Y_e$ on the $r-\dM$ plane. 
(a) $\alpha=0.01$ and $a=0$. (b) $\alpha=0.01$ and $a=0.95$. 
(c) $\alpha=0.1$ and $a=0$.  (d) $\alpha=0.1$ and $a=0.95$.
} 
\end{center}
\end{figure} 

\begin{figure}
\begin{center}
\plotone{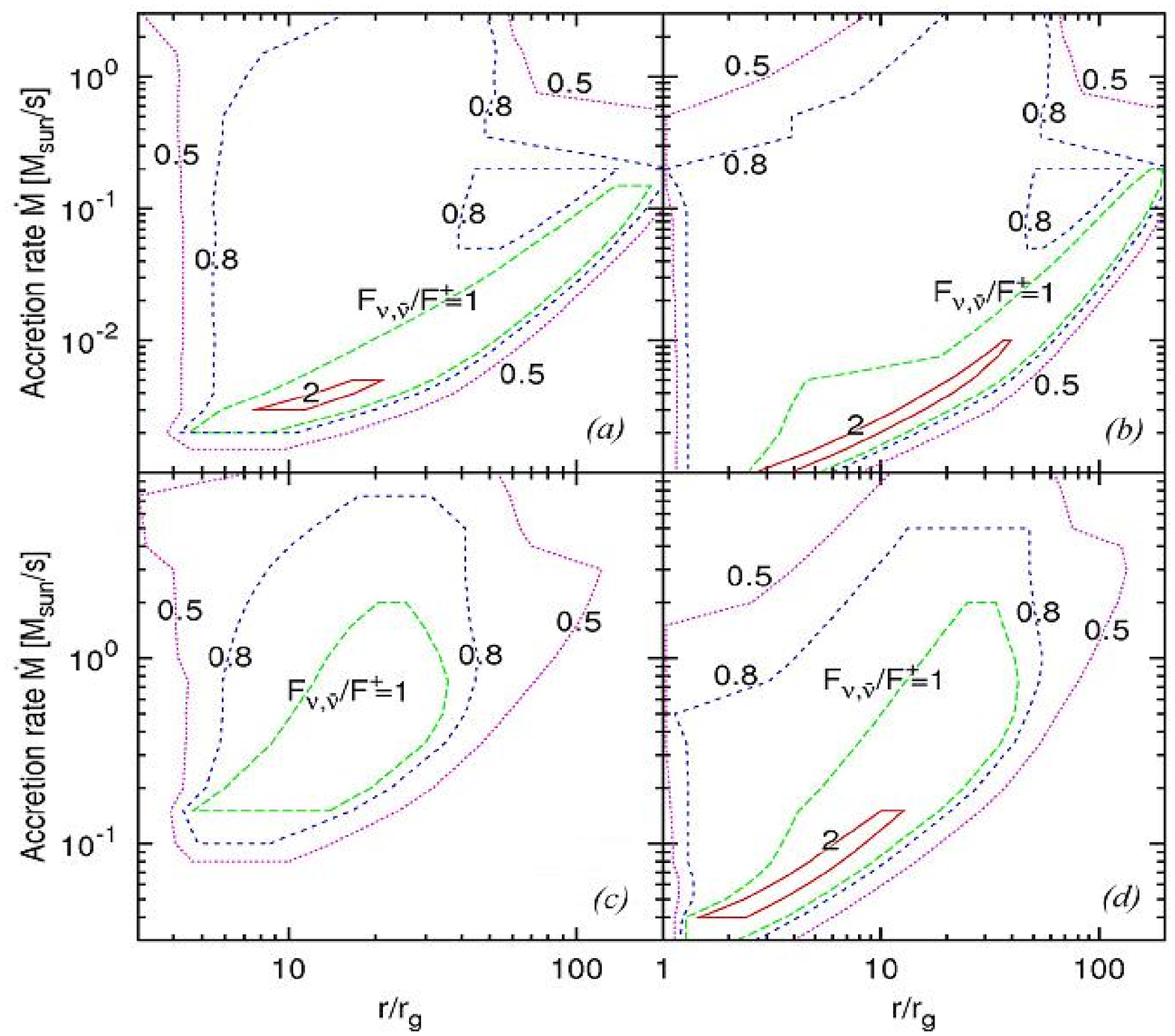}
\caption{
Contours of local cooling efficiency $(F_\nu+F_{\bar\nu})/F^+$ on the 
$r-\dM$ plane. 
(a) $\alpha=0.01$ and $a=0$. (b) $\alpha=0.01$ and $a=0.95$. 
(c) $\alpha=0.1$ and $a=0$.  (d) $\alpha=0.1$ and $a=0.95$.
} 
\end{center}
\end{figure} 

It is worth emphasizing that all three characteristic accretion rates,
$\dMign$, $\dMop$, and $\dMtr$ are lower for disks with smaller 
viscosity parameter $\alpha$. The low-$\alpha$ disks are denser and
have significantly larger $r_\nu$ and $\rtr$.
For example, the radius of opaqueness $r_\nu$ in the models with 
$\alpha=0.01$ is almost one order of magnitude larger than in 
$\alpha=0.1$ models. 

The dependence of $\dMign$, $\dMop$, and $\dMtr$ on $\alpha$ is well 
approximated by a power law (see Fig.~13),
\be
\label{eq:pw}
   \dMign=\Kign\left(\frac{\alpha}{0.1}\right)^{5/3}, \qquad 
   \dMop=\Kop\left(\frac{\alpha}{0.1}\right), \qquad 
   \dMtr=\Ktr\left(\frac{\alpha}{0.1}\right)^{1/3}.
\ee
where the normalization factors $K$ depend on the black hole spin $a$.
For $a=0$, we find 
 $\Kign=0.071M_\odot$~s$^{-1}$, $\Kop=0.7M_\odot$~s$^{-1}$, and 
 $\Ktr=9.3M_\odot$~s$^{-1}$.
For $a=0.95$, we find 
 $\Kign=0.021M_\odot$~s$^{-1}$,
 $\Kop=0.06M_\odot$~s$^{-1}$, and $\Ktr=1.8M_\odot$~s$^{-1}$.

\begin{figure}
\begin{center}
\plottwo{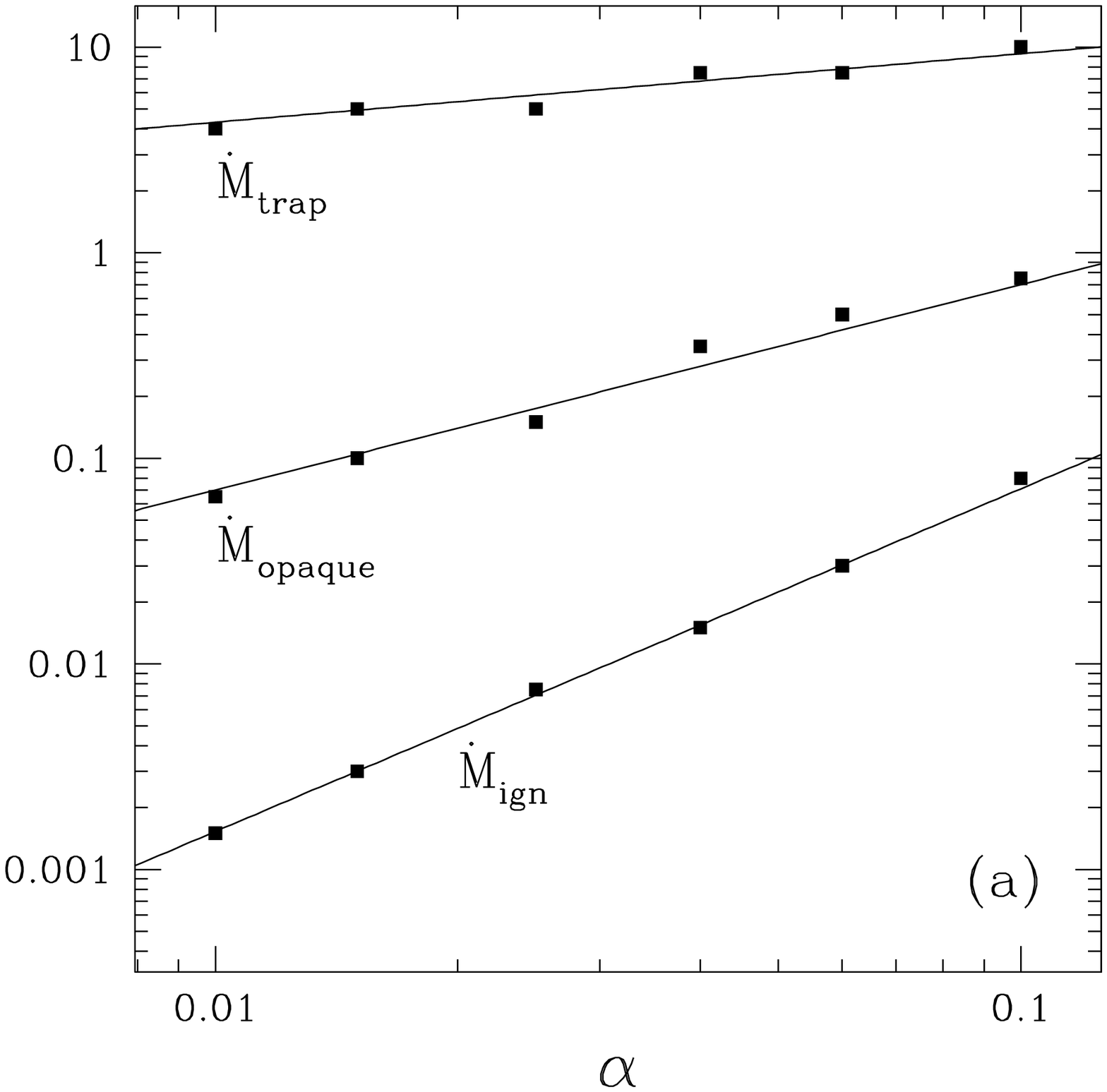}{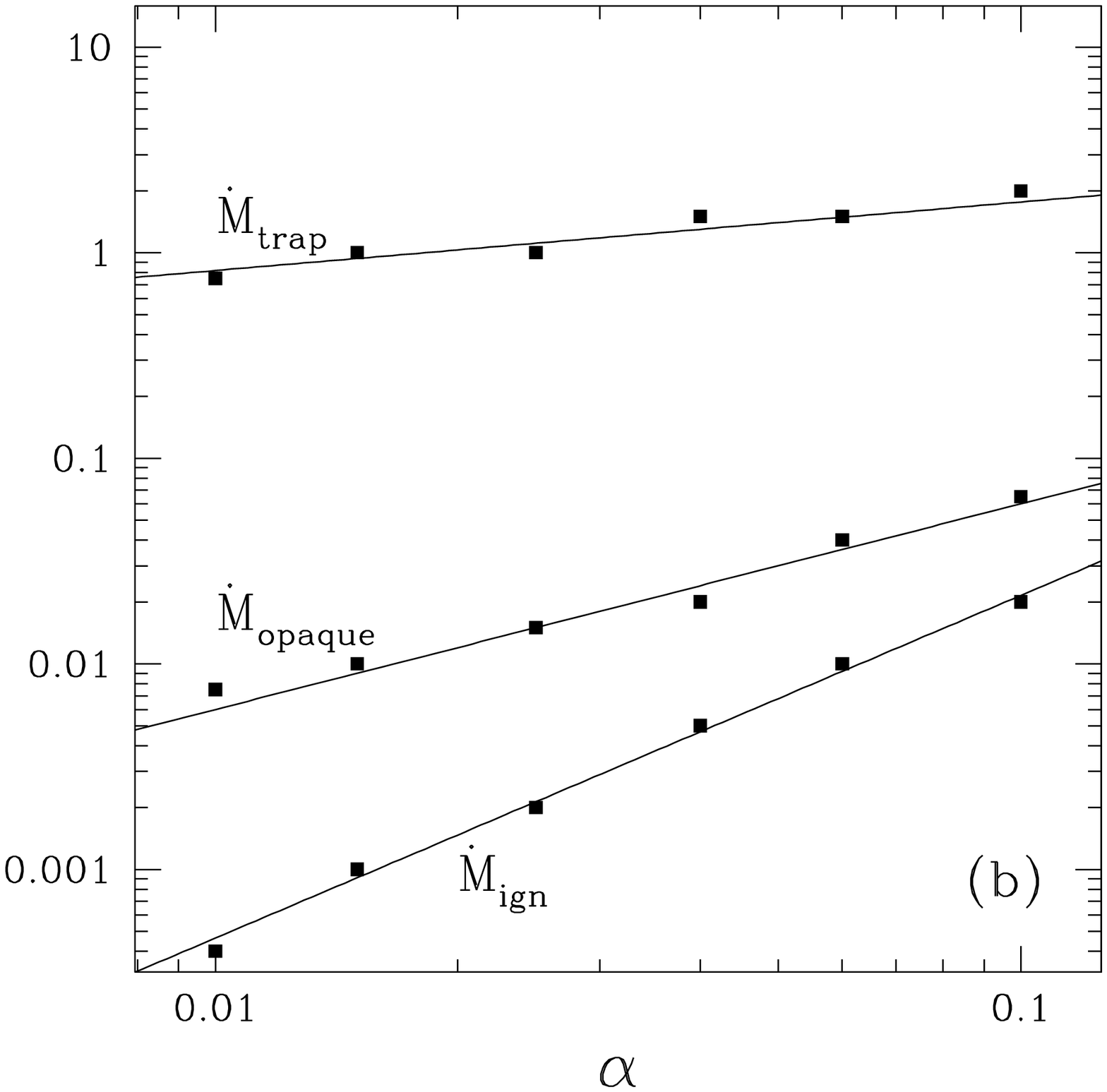}
\caption{
Characteristic accretion rates $\dMign$, $\dMop$, and $\dMtr$.
The lines show the power-law approximation (eq.~\ref{eq:pw}).
(a) Schwarzschild disk ($a=0$). (b) Kerr disk ($a=0.95$).
} 
\end{center}
\end{figure} 

The maximum radiative efficiency of the disk, $L_\nu/\dM c^2$, is determined
by the binding energy at the last stable orbit $\rms$. It equals 
0.057 for $a=0$ and $0.19$ for $a=0.95$. The real efficiency is 
somewhat smaller because part of the released energy remains stored
in the disk and advected into the black hole. The efficiency is shown
as a function of $\dM$ in Figure~18.

\begin{figure}
\begin{center}
\plotone{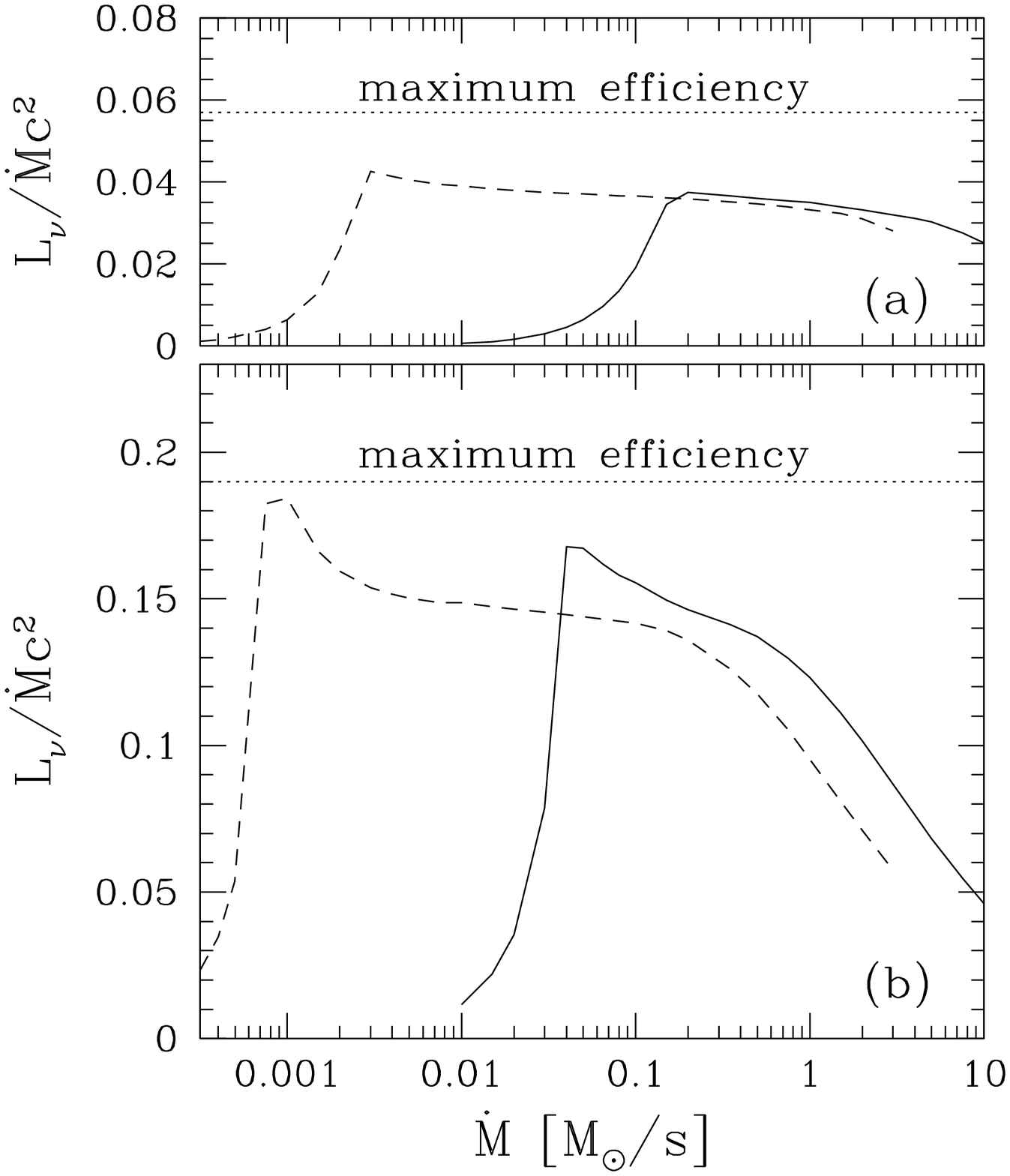}
\caption{
Net efficiency of neutrino cooling of the disk $L_\nu/\dM c^2$. The cases 
of $\alpha=0.1$ and $\alpha=0.01$ are shown by the solid and dashed
curves, respectively. The maximum efficiency that corresponds to complete 
cooling is shown by the horizontal dotted curve. (a) Schwarzschild disk
($a=0$). (b) Kerr disk ($a=0.95$).
} 
\end{center}
\end{figure} 


\section{CONCLUSIONS}

This paper presents the model of neutrino-cooled accretion disks 
around rotating black holes that calculates self-consistently nuclear 
composition, neutrino emission, and fluid dynamics in the Kerr metric. 
The model is one-dimensional in the sense that all parameters of the disk 
are integrated/averaged in the vertical direction and depend on $r$ only, 
and the standard $\alpha$-prescription is used for viscosity.
The model also assumes a steady state, which is applicable only to 
radii where accretion timescale is shorter than the timescale of 
variation of $\dot{M}$ in the problem.
The main advance of our work compared with PWF is the calculation of 
electron degeneracy and nuclear composition of the accreting matter; 
both dramatically affect the disk and its neutrino emission.

The disk has a clear structure with five characteristic
radii $r_\alpha$, $\rign$, $r_\nu$, $r_{\bar\nu}$, and $\rtr$.
Radial advection of lepton number $Y_e$ and viscously dissipated heat 
is important outside the ignition radius $\rign$. Most of the viscous 
heat is lost to neutrino emission at $r<\rign$.

The neutrino-cooled disk forms at accretion rates $\dM>\dMign$ 
which depends on the black-hole spin $a$ and the viscosity parameter 
$\alpha$ (see Fig.~17 and eq.~\ref{eq:pw}).
General properties of the $\nu$-cooled disk are as follows.

1.  --- The disk is relatively thin, $H/r\sim 0.1-0.3$, 
especially in the inner region where most of accretion energy is released.
The outer advective region $r\simgt 100r_g$ is also significantly cooled 
by the gradual disintegration of $\alpha$-particles, and its thickness is 
reduced. 

2. --- The $\nu$-cooled disk is nearly in 
$\beta$-equilibrium, in agreement with analytical 
estimates (Beloborodov 2003a). In particular, the relation between 
$\rho$, $T$, and $Y_e$ calculated under the equilibrium assumption
(Fig.~1 and 2 in Beloborodov 2003a) applies with a high accuracy.

3. --- Degeneracy of electrons in the disk significantly suppresses 
the positron density $n_{e^+}$. However, the strong degeneracy limit is 
not applicable --- the disk regulates itself to a mildly degenerate state.
The reason of this regulation is the negative feedback of degeneracy on 
the cooling rate: higher degeneracy $\mu_e /kT$ $\rightarrow$ fewer 
electrons (lower $Y_e$) and positrons ($n_{e^+}/n_{e^-}\sim e^{-\mu_e/kT}$) 
$\rightarrow$ weaker neutrino emission $\rightarrow$ 
lower cooling rate $\rightarrow$ higher temperature $\rightarrow$ lower 
degeneracy. 

4. --- Pressure in $\nu$-cooled disks is dominated by baryons,
$P\approx P_b=(\rho/m_p)kT$, most of which are neutrons.

5. --- All $\nu$-cooled disks are very neutron rich in the inner region, 
with $Y_e\sim 0.1$ or lower.

The high neutron richness has important 
implications for the global picture of GRB explosion (e.g. Derishev, 
Kocharovsky, \& Kocharovsky 1999; Beloborodov 2003a,b; Rossi, Beloborodov, 
\& Rees 2006). When the neutron-rich material is ejected in a 
relativistic jet, it develops a large Lorentz factor, and the neutrons 
gradually decay 
on scales up to $10^{17}$~cm where the GRB blast wave is observed.

The disks around rapidly rotating black holes are markedly different from 
disks around Schwarzschild black holes. They extend much closer to the 
center and reach higher temperatures and densities. For example, we found 
that the disk with $\alpha=0.1$ around a Kerr black hole with $a=0.95$ 
becomes opaque for neutrinos at $\dMop\sim 0.07M_\odot$~s$^{-1}$, which is 
10 times lower than the corresponding $\dMop$ for a Schwarzschild black hole. 
Besides, Kerr disks produce much higher neutrino fluxes in the inner
region, with a higher mean energy per neutrino. The annihilation reaction
$\nu+\bar{\nu}\rightarrow e^++e^-$ deposits energy above the disk 
and can drive a powerful outflow that will be investigated elsewhere.

\acknowledgements
This work was supported by NASA grant NAG5-13382.


\section*{Appendix A: Radial Velocity in a Relativistic Disk}
                                                                     
Conservation of energy and angular momentum is expressed by the
equations (see Beloborodov, Abramowicz, \& Novikov 1997)
\be 
\label{eq:1}
   \frac{d}{dr} \left[ \mu\left( \frac{\dot M u_t}{2\pi}+
     2\nu\Sigma r \sigma^r_{\; t} \right) \right] =\frac{F^{-}}{c^2}ru_t, 
\ee
\be 
\label{eq:2}
  \frac{d}{dr} \left[ \mu\left( \frac{\dot M u_\phi}{2\pi}+2\nu\Sigma
    r\sigma^r_{\;\phi} \right) \right] = \frac{F^{-}}{c^2}ru_\phi,
\ee
where 
\be
\label{eq:sigma}
  \sigma^r_{\;\phi}=\frac{1}{2}g^{rr}g_{\phi\phi}\sqrt{-g^{tt}} 
  \gamma^3 \frac{d\Omega}{dr}
\ee
is the shear and $\mu=(U+P)/\rho c^2$ is the relativistic enthalpy per 
unit rest mass ($\mu\approx 1$ in the neutrino-cooled disk); 
$\gamma=u^t/\sqrt{-g^{tt}}$ and $\Omega=u^\phi/u^t$ 
are taken for the Keplerian circular motion (eq.~\ref{eq:Omega}).
From equations~(\ref{eq:1}) and (\ref{eq:2}) one can derive (see 
also Page \& Thorne 1974) 
\begin{eqnarray*}  
  2   \nu\Sigma r\sigma^r_{\;\phi} = T(x) & = &  
   -\frac{    \dot M}{2\pi}\frac{GM}{c}\frac{x^3+a}{(x^3-3x+2a)^{1/2}x^{3/2}}
       \left[ (x-x_0)-\frac{3}{2}a\ln(\frac{x}{x_0})\right.  \\
    &{}&-\frac{3(x_1-a)^2}{x_1(x_1-x_2)(x_1-x_3)}\ln(\frac{x-x_1}{x_0-x_1})
       -\frac{3(x_2-a)^2}{x_2(x_2-x_1)(x_2-x_3)}\ln(\frac{x-x_2}{x_0-x_2}) \\
    &{}&\left. -\frac{3(x_3-a)^2}{x_3(x_3-x_1)(x_3-x_2)}\ln(\frac{x-x_3}
       {x_0-x_3})\right]
\label{eq:T}
\end{eqnarray*}
where $x=(rc^2/GM)^{1/2}$, $x_1,x_2,x_3$ are the three roots of equation
$x^3-3x+2a=0$, and $x_0$ corresponds to the marginally stable orbit
$\rms$ where fluid falls freely into the black hole and zero viscous 
torque is assumed.

The radial velocity $u^r$ may now be expressed as
\be  
   u^r=\frac{\dot M}{2\pi r \Sigma}
      =\frac{\dot M}{\pi}\,\frac{\nu\, \sigma^r_{\;\phi}}{T} 
\ee
where $\sigma^r_{\;\phi}(r)$ and $T(r)$ are known functions 
given by equations~(\ref{eq:sigma}) and (\ref{eq:T}).
Substituting the $\alpha$ prescription for the kinematic viscosity
coefficient, $\nu=(2/3)\alpha c_s H$, one finds
\be
    u^r =\alpha\, c_s \left(\frac{H}{r}\right) \frac{2\dot M}{3\pi}
           \frac{ r\sigma^r_{\;\phi}}{T}
        = \alpha\, c_s \left(\frac{H}{r}\right) S^{-1}(r).
\ee
The numerical factor $S(r)=(3\pi/2)(T/\dM r\sigma^r_{\;\phi})$ varies from 
zero at the inner radius $\rms$ to unity at $r\gg r_g$. 

``Newtonian'' approximation that is often used in the literature on
accretion disks (including GRB disks) is given by
$$
  S_{\rm N}(r)=1-\left(\frac{\rms}{r}\right)^{1/2}.
$$
It is derived for the accretion disk in Newtonian space by requiring
conservation of angular momentum and imposing zero torque at a 
specified radius $\rms$, e.g. $\rms=3r_g$ to mimic a Schwarzschild 
spacetime (Shakura \& Sunyaev 1973). The correct function $S$ 
differs significantly from $S_N$ even for a Schwarzschild black hole: 
$S\simlt S_N/2$ in the inner region of the disk.


\section*{Appendix B: Cross Sections for Neutrino Interactions}

We summarize here the cross sections of neutrino reactions that we use
in this paper (see Burrows \& Thompson 2002 for a recent review of the 
reactions). The cross sections are expressed in terms of $\sigma_0$,
\be 
 \sigma_0=\frac{4G_F^2(m_ec^2)^2}{\pi (\hbar c)^4}
        \simeq1.71\times 10^{-44}\textmd{~cm}^2.   
\ee
The neutrino energy is denoted by $E$ and expressed in units of $m_e c^2$. 
                                                                           
1. --- Neutrino absorption by nucleons:
\be  
   \nu + n \rightarrow e^{-} + p,  \hspace{3pc}  
   \bar\nu + p \rightarrow e^{+} + n 
\ee
The cross section of $\nu$ absorption by neutron is given by
(e.g. Bemporad et al. 2002)
\be  
   \sigma_{\nu n}(E_\nu)=\sigma_0 \left( \frac{1+3g_A^2}{4} \right)
                     (E_\nu+Q)^2\sqrt{1-\frac{1}{(E_\nu+Q)^2}},
\ee
where 
$Q=(m_n-m_p)/m_e=2.53$, $g_A\simeq -1.26$ is the axial coupling constant, 
$(1+3g_A^2)/4\simeq 1.44$. This approximation neglects the recoil of neutron, 
however, it has only a small error $<1.5$\% when the neutrino energy is 
below 80~MeV (Strumia \& Vissani 2003). For $\bar\nu$ absorption by
protons, however, there is a significant correction due to the recoil,
and a better approximation should be used. We use the approximation of
Strumia \& Vissani (2003),
\begin{eqnarray*}
   \sigma_{\bar\nu p}(E_{\bar\nu})
          &=&10^{-43}\,\kappa^2 (E_{\bar\nu}-Q)^2
             \sqrt{1-\frac{1}{\kappa^2(E_{\bar\nu}-Q)^2}} \\
          &{}&\times(\kappa E_{\bar\nu})^{-0.07056+0.02018
              \ln (\kappa E_{\bar\nu})-0.001953\ln^3 (\kappa E_{\bar\nu})}
              {\rm ~cm}^2.
\end{eqnarray*}
where $\kappa=0.511$.  \\
\medskip
                                                                                
2. --- Neutrino-baryon elastic scattering.
The cross sections of scattering on proton, neutron,
and $\alpha$ particles are
\be  
   \sigma_p(E)=\frac{\sigma_0 E^2}{4}
   \left[ 4\sin^4\theta_W-2\sin^2\theta_W+\frac{1+3g_A^2}{4} \right]
       =0.30\,\sigma_0 E^2,  
\ee
\be  
   \sigma_n(E)=\frac{\sigma_0 E^2}{4}\left( \frac{1+3g_A^2}{4} \right)
          =0.36\,\sigma_0 E^2, 
\ee
\be 
  \sigma_\alpha (E)=4\,\sigma_0\,\sin^4\theta_W E^2=0.21\,\sigma_0 E^2,    
\ee
where $\theta_W$ is the Weinberg angle and $\sin^2\theta_W=0.23$.
\medskip
                                                                                
3. --- Neutrino-electron (or neutrino-positron) scattering.
The cross section of scattering for $\nu$ or $\bar\nu$ is given by
(Burrows \& Thompson 2002),
\be  
  \sigma_e(E)=\frac{3}{8}\sigma_0\,\theta\,E\,\left(1+\frac{\eta_e}{4}\right)
                         \left[(C_V+C_A)^2+\frac{1}{3}(C_V-C_A)^2\right],
\ee
where 
$\theta=kT/(m_ec^2)$ is temperature, $\eta_e=\mu_e/kT$ is the degeneracy
parameter, $C_A=+1/2$ for $\nu$ and $C_A=-1/2$ for $\bar\nu_e$, and 
$C_V=\frac{1}{2}+2\sin^2\theta_W$ for both $\nu$ and $\bar\nu$.
                                                                                
4. --- Neutrino annihilation: $\nu+\bar\nu\rightarrow e^-+e^+$.
Considering both the neutral and charged current reactions, the total 
cross section at high energies $E_\nu,E_{\bar\nu}\gg 1$ is given by
\be  
  \sigma_{\nu\bar\nu}=K_{\nu\bar\nu}\sigma_0
    \frac{(\mathbf{P_\nu}\cdot \mathbf{P_{\bar\nu}})^2}{E_\nu E_{\bar\nu}}, 
\ee
where $\mathbf{P}_\nu$ and $\mathbf{P}_{\bar\nu}$ are the four-momenta of 
$\nu$ and $\bar\nu$ in units of $m_e c$, and $K_{\nu\bar\nu}=
(1+4\sin^2\theta_W+8\sin^4\theta_W)/12=0.195$ (Goodman et al. 1987; 
the full expression is found in Dicus 1972). After averaging over 
target distribution, one gets the average cross section for neutrino 
and anti-neutrino,
\be  
  \sigma_\nu(E_\nu)=\frac{4}{3}K_{\nu\bar\nu}\sigma_0 E_\nu \bar E_{\bar\nu} 
\ee
\be  
  \sigma_{\bar\nu}(E_{\bar\nu})
  =\frac{4}{3}K_{\nu\bar\nu}\sigma_0 E_{\bar\nu}\bar E_\nu  
\ee
where $\bar E_\nu$ and $\bar{E}_{\bar\nu}$ are the average energies 
of neutrinos and anti-neutrinos, respectively.



\begin{references}  

\reference{}
Balbus, S. A., \& Hawley J. F. 1998, Rev. Mod. Phys., 70, 1

\reference{}
Beloborodov, A. M. 1998, MNRAS, 297, 739

\reference{}
Beloborodov, A. M. 1999, ASP Conference Series 161, 295

\reference{}
Beloborodov, A. M. 2003a, ApJ, 588, 931 

\reference{}
Beloborodov, A. M. 2003b, ApJ, 585, L19 

\reference{}
Beloborodov, A. M., Abramowicz, M. A., \& Novikov, I. D. 1997, ApJ, 491, 267 

\reference{}
Bemporad, C., Gratta, G., \& Vogel, P. 2002, Rev. Mod. Phys, 74, 297 

\reference{}
Burrows, A., \& Thompson, T.A. 2002, astro-ph/0211404  

\reference{}
Dicus, D. A. 1972, Phys. Rev. D, 6, 941 

\reference{}
Derishev, E. V., Kocharovsky, V. V., \& Kocharovsky, Vl. V. 1999, ApJ, 521, 640

\reference{}
Di Matteo, T., Perna, R., \& Narayan, R. 2002, ApJ, 579, 706 

\reference{}
Goodman, J., Dar, A., \& Nussinov, S. 1987, ApJ, 314, L7 

\reference{}
Kohri, K., \& Mineshige, S. 2002, ApJ, 577, 311 

\reference{}
Kohri, K., Narayan, R.\& Piran, T. 2005, ApJ, 629, 341 

\reference{}
MacFadyen, A. I., \& Woosley, S.E. 1999, ApJ, 524, 262 

\reference{}
Meyer, B. S. 1994, ARA\&A, 32, 153

\reference{}
Misner, C. W., Thorne, K. S., \& Wheeler, J. A. 1973, Gravitation 
(San Francisco: Freeman)

\reference{}
Narayan, R., Piran, T., \& Kumar, P. 2001, ApJ, 557, 949

\reference{}
Narayan, R., \& Yi, I. 1994, ApJ, L13

\reference{}
Paczy{\'n}ski, B. 1978, AcA, 28, 91

\reference{}
Page, D. N., \& Thorne, K. S. 1974, ApJ, 191, 499 

\reference{}
Piran, T. 2004, Rev. Mod. Phys, 76, 1143 

\reference{}
Popham, R., Woosley, S. E., \& Fryer, C. 1999, ApJ, 518, 356 (PWF)

\reference{}
Pruet, J., Woosley, S. E., \& Hoffman, R. D. 2003, 586, 1254

\reference{}
Rossi, E.,  Beloborodov, A. M., \& Rees, M. J. 2006, MNRAS, in press
(astro-ph/0512495)

\reference{}
Ruffert, M., Janka, H.-T., Takahashi, K., \& Schaefer, G. 1997, A\&A, 319, 122

\reference{}
Sawyer, R. F. 2003, Phys. Rev. D, 68, 063001

\reference{}
Shakura, N. I., \& Sunyaev, R. A. 1973, A\&A, 24, 337

\reference{}
Shapiro, S. L., \& Teukolsky, S. L. 1983, Black Holes, White Dwarfs, 
and Neutron Stars (New York: Wiley)

\reference{}
Strumia, A., \& Vissani, F. 2003, Phys. Lett. B, 564, 42 

\reference{}
Thompson, T. A., Burrows, A., Horvath 2000, Phys., C62, 35802

\reference{}
Woosley, S. E. 1993, ApJ, 405, 273

\end{references}
\end{document}